\documentclass[prx,reprint,amsmath,amssymb,superscriptaddress,floatfix,longbibliography]{revtex4-1}
\usepackage[breaklinks=true,colorlinks,citecolor=blue,linkcolor=blue,urlcolor=blue]{hyperref}
\usepackage{graphicx,epsfig}
\usepackage{amsmath}
\usepackage{xcolor,color}
\usepackage{subfigure}
\usepackage{color}

\def\be{\begin{equation}}
\def\ee{\end{equation}}
\def\nn{\nonumber}
\def\bea{\begin{eqnarray}}
\def\eea{\end{eqnarray}}

\newcommand\inv[1]{#1\raisebox{1.15ex}{$\scriptscriptstyle-\!1$}}

\begin{document}
\title{Linear magnetochiral transport in tilted type-I and type-II Weyl semimetals}
\author{Kamal Das}
\email{kamaldas@iitk.ac.in}
\affiliation{Dept. of Physics, Indian Institute of Technology Kanpur, Kanpur 208016, India}
\author{Amit Agarwal}
\email{amitag@iitk.com}
\affiliation{Dept. of Physics, Indian Institute of Technology Kanpur, Kanpur 208016, India}

\begin{abstract}
Berry curvature in Weyl semimetals leads to intriguing magnetoconductivity and magneto-thermal transport properties.
Here, we explore the impact of the tilting of the Weyl nodes, on the magnetoconductivity of  type-I and type-II Weyl  semimetals using the Berry curvature connected Boltzmann transport formalism. We find that in addition to the quadratic magnetic field ($B$) corrections induced by the tilt, there are also anisotropic and $B$-linear corrections in several elements of the conductivity matrix. For the case of magnetic field applied perpendicular to the tilt direction, we show the existence of previously unexplored $B$-linear transverse conductivity components. For the other case of magnetic field applied parallel to the tilt axis, the $B$-linear corrections appear in the longitudinal conductivity 
giving rise to anisotropic magnetoresistance measurements. 
Our systematic analysis of the full magnetoconductivity matrix, predicts several specific experimental signatures related to the tilting of the Weyl nodes in both type-I and type-II Weyl semimetals.
\end{abstract}
\maketitle

\section{introduction}

Topological phases of matter are of immense interest owing to their new, fundamental, and exciting physics, along with the promise of potential applications \cite{Hasan2010,kong11,Bansil2016,Ashvin2018, Claudia_review2017}. Amongst the topological semimetal phases, the Weyl semimetal (WSM) phase, hosting 
linearly dispersing quasiparticles with a distinct chirality, is very interesting and has been discovered experimentally in three dimensional (3D) condensed matter systems \cite{Kim2013,xu2015a,Huang2015,Lv2015,xu2015b}. The WSM phase  is 
very robust, requiring only the translational invariance of the crystal \cite{Ashvin2011,burkov_balents2011}. 
In any topological lattice theory, which is invariant under the gauge field and the action is bilinear in the fermion fields, the chiral Weyl fermions 
always come in pairs of opposite chirality \cite{Nielsen1981}. Depending on the broken inversion or time reversal symmetry (TRS), the two Weyl nodes (WNs) of opposite chirality 
are separated either in energy or momentum. Unlike their high-energy counterparts, the WNs in condensed matter systems can also be anisotropic and tilted 
(called type-I) or tilted over (type-II) \cite{Soluyanov2015, Pengli2017, Yazyev2016, Jiang2017, Pal2018, Zhang2018, Chang2016, Xu2017}, as shown in Fig.~\ref{f1}. In type-I WSM, there is a closed Fermi surface enclosing either an electron or a hole pocket, with the vanishing density of states at the Weyl point. 
However, in type-II WSM, there are unbounded electron and hole pockets at the Fermi surface, and there are a large density of states even at the Weyl point {\cite{Soluyanov2015}, 
which results in different magnetotransport properties of type-II WSM \cite{Saha2018, Zyuzin_A_A2016, G_Sharma2017a, Mukherjee2018, Nandini2017, Ferreiros2017}.

The WSM phase was first proposed in the magnetic pyrochlore iridates \cite{Ashvin2011,Xu2011,Bulmash2014} which are TRS broken systems. The TRS broken WSM phase can also be induced by applying a magnetic field in 3D-Dirac semimetals like Na$_{3}$Bi, Cd$_{3}$As$_{2}$, Bi${}_{1-x}$Sb${}_{x}$ \cite{Li2016, Xiong2015, Desrat2015,Kim2013}. The inversion symmetry broken WSM phase has been reported in transition metal mononictides (TX, T=Ta/Nb,  X=As/P) \cite{Weng2015,Lv2015,Sudesh2017,Niemann2017,Hu2016,Arnold2016,Li2017,Huang2015,Liang18}.  The type-II WSM phase has been demonstrated in WTe${}_{2}$ \cite{Pengli2017}, MoTe${}_{2}$ \cite{Jiang2017}, and YbMnBi${}_{2}$ \cite{Pal2018}, based on angle resolved photoemission spectroscopy, electronic transport, and optical experiments. 

Weyl nodes act as a sink or source of the Berry curvature (BC) \cite{Berry1984}, which in turn acts as a magnetic field in the momentum space \cite{Xiao2010}. This has a significant impact on the dynamics of the electron wave-packets in the WSM lattice, particularly in the presence of parallel electric and magnetic fields (a finite ${\bf E}\cdot{\bf B}$ term). This results in a plethora of interesting magnetoconductivity (MC) and magneto-thermal transport properties in WSM, which have been extensively investigated theoretically \cite{Nielsen1983,PhysRevB.86.115133,Son12,Son_Spivak2013,Sinitsyn2008,Haldane2004,PhysRevB.89.075124,kim_kim,PhysRevB.90.165115,PhysRevB.91.214405, 
PhysRevB.91.245157,PhysRevB.93.155125,PhysRevB.95.165135,Burkov2017,Ferreiros2017,PhysRevB.96.155138,PhysRevB.96.235134,Nandini2017,Nandy2017,Burkov2017,PhysRevB.86.115133,G_Sharma2017b, G_Sharma2016,Wei18}. For example, Negative magnetoresistance (MR) has been observed in several WSMs, including all the members of the TaAs family \cite{Sudesh2017,Niemann2017,Hu2016,Arnold2016,Li2017} and WSMs derived from 3D-Dirac semimetals \cite{Li2016, Xiong2015, Desrat2015, Kim2013}, among others. 
The chiral magnetic effect \cite{Li16} and the anomalous Hall effect have been reported in ZrTe${}_{5}$ \cite{Liang2018}. The planar Hall effect 
has been demonstrated in GdPtBi \cite{Nitesh2018} and TaP \cite{Yang18}. 
Additionally the anomalous Nernst effect has been demonstrated in Cd$_3$As$_2$ \cite{Liang2017}, along with other thermoelectric effects \cite{Jia2016}. 

\begin{figure}[t]
\includegraphics[width = \linewidth]{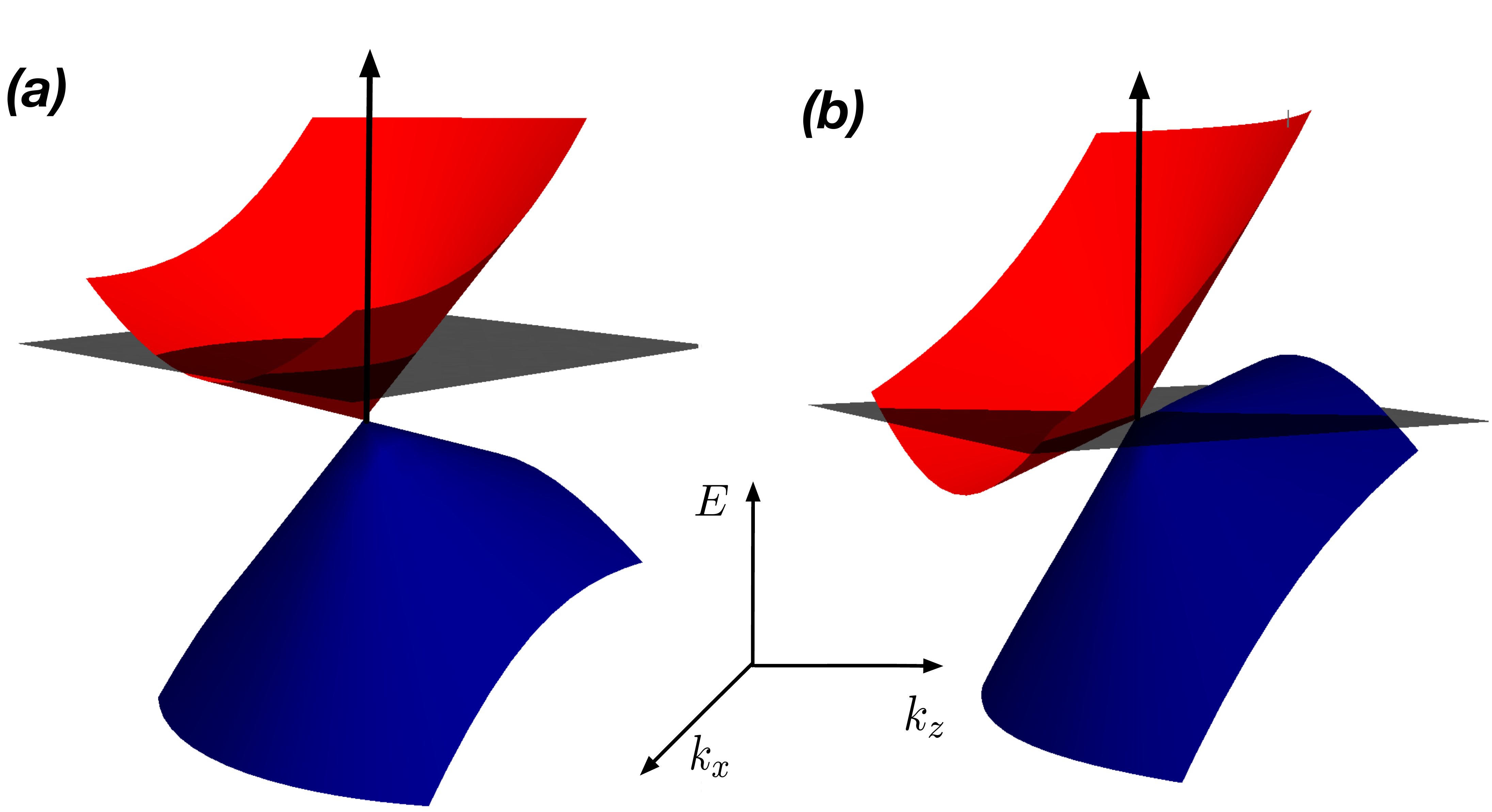}
\caption{(a) Schematic of a tilted type-I WSM showing a single Weyl node with a closed Fermi surface and an electron pocket (marked in red) at the Fermi energy. 
(b) Schematic of a ``tilter over" type-II WSM node with an unbounded Fermi surface having an electron (red) and a hole pocket (blue) even at the Weyl point.  
\label{f1}}
\end{figure}

While there are several theoretical works focused on different aspects of MC in WSM, a systematic analysis of the full conductivity matrix, covering all possible cases for tilted type-I and type-II WSM is still lacking. This is the focus of this paper. Our primary aim is to uncover the tilt and the magnetic field dependence of various components of the MC, using the semi-classical BC connected Boltzmann transport formalism.  
For tilted Weyl nodes (both type-I and type-II), we predict tilt dependent $B$-linear corrections to the longitudinal (when ${\bf B}$ is along the tilt axis) as well as the transverse MC (when ${\bf B}$ is perpendicular to  the tilt axis). For these $B$-linear corrections, the corresponding magneto-current is found to either ${\bf j} \propto ({\bf E}\cdot{\bf B}) \hat{\bf R}$, or  ${\bf j} \propto ({\bf E}\cdot\hat{\bf R}){\bf B}$, or ${\bf j} \propto ({\bf B}\cdot\hat{\bf R}) {\bf E}$, where $\hat{\bf R}$ denotes the direction of the tilt axis.  Additionally, we also find tilt dependent quadratic-$B$ correction to the MC along all directions in both type-I and type-II WSM.
Together, the $B$-linear and the quadratic-$B$ terms combine result in anisotropic longitudinal magnetoresistance \cite{Pengli2017}, defined as ${\rm MR} = \sigma_{ii}(0)/\sigma_{ii}(B) - 1$ in both type-I and type-II WSM, whose sign depends on the direction of the applied magnetic field. Finally, our results also generalize the known results for longitudinal and the planar Hall conductivity to include the tilt dependence.

The paper is organized as follows: We start with the review of the BC-connected Boltzmann transport formalism in Sec. \ref{review}. This is followed by a systematic discussion of the MC matrix for the three different cases of the isotropic WSM, type-I WSM, and type-II WSM which are presented in Secs.~\ref{Iso}, ~\ref{type-I}, and \ref{type-II}, respectively. 
We specifically consider the different cases of magnetic fields being along the tilt direction, or perpendicular to it, and allow for the possibility of the two Weyl nodes to have different tilts angles.  Anisotropic MR is discussed in Sec.~\ref{MR} along with the limiting cases in Sec.~\ref{limiting}. 
Finally, we summarize our findings in Sec.~\ref{Concl}.

\section{Berry-curvature-connected Boltzmann transport formalism}\label{review}
We begin by reviewing the Boltzmann transport formalism, with a focus on understanding the effect of applied magnetic and electric fields on the charge transport properties of materials with finite BC. We will use this to obtain the anisotropic and linear magnetotransport properties of tilted Weyl nodes. The semiclassical Boltzmann transport approach works well for small magnetic fields and small cyclotron frequency $\omega_c$, where the Landau quantization can be ignored \cite{Jia2016}. It is valid in the regime $\hbar \omega_c \ll \mu$, with $\mu$ denoting the chemical potential. 
Within the linear response theory, the phenomenological transport equation for the electrical current ${\bf j}^{e}$ is given by \cite{ashcroft}, 
\begin{eqnarray}\label{tetc}
j_{i}^{e}=\sigma_{ij}E_{j}~.
\end{eqnarray}
Here, $i$ and $j$ are spatial coordinate indices (running over $x$, $y$, and $z$), $E_j$ denotes the external electric field along the $j$th coordinate, and $\sigma_{ij}$ denote the elements of the electrical conductivity matrix.

In the Boltzmann transport formalism, the conductivity matrix is calculated by doing a Brillouin zone (BZ) sum over the relevant physical quantity (velocity operator) keeping only the physically occupied states. This explicitly requires three things, all of which are influenced by the presence of a finite BC: 
(1) the equation of motion (EOM) describing the dynamics of the center of the carrier wave-packet in a given band (in terms of the center position and the corresponding Bloch wave-vector),  (2) the non-equilibrium distribution (NDF) function specifying the occupancy of the bands under external perturbation, and (3) the phase-space volume which gets modified in presence of finite BC and an external magnetic field. 

\subsection{Berry curvature dependence of the ``three elements"}
The EOM for the carrier (center of mass of the wave-packet) location $\bf r$ and the corresponding Bloch wave vector $\bf k$ in a given band is given by \cite{sundaram,son_yamamoto}
\bea \label{EOM1}
\dot{\bf r} &=& \frac{1}{\hbar}\nabla_{\bf k}{\epsilon}_{\bf k} -  \dot{\bf k} \times {\bf \Omega}_{\bf k}~, \\
\hbar \dot{\bf k} &=& -e {\bf E} -e \dot{\bf r} \times {\bf B}~, \label{EOM2}
\eea
where `$-e$' is the electronic charge and ${ \epsilon}_{\bf k}$ is the electronic dispersion. The BC is given by $ {\bf \Omega}_{\bf k} = \nabla_{\bf k} \times {\bf A}_{{\bf k}}$, where ${\bf A}_{{\bf k}} = i \langle u_{\bf k}| \nabla_{\bf k} u_{\bf k}\rangle$   and $|u_{\bf{k}}\rangle$ is the Bloch wave function. A finite BC acts as a ``fictitious magnetic field" in the reciprocal space, as evidenced by the second term on the right hand side of Eq.~\eqref{EOM1}. 

Equations~\eqref{EOM1} and \eqref{EOM2} can be decoupled, to obtain \cite{Son_Spivak2013} 
\begin{eqnarray}\label{eom_r}
\dot{\bf r}& = &D_{\bf k}\left[{{\bf v}}_{\bf{k}}+\frac{e}{\hbar}({\bf E}\times {\bf \Omega}_{\bf k})+\frac{e}{\hbar}({{\bf v}}_{\bf{k}}\cdot {\bf \Omega}_{\bf k}){\bf B}\right],\\\label{eom_k}
\hbar\dot{\bf k }&=& D_{\bf k}\left[-e{\bf E} - e({{\bf v}}_{\bf{k}}\times {\bf B})-\frac{e^{2}}{\hbar}({\bf E}\cdot{\bf B}){\bf  \Omega_{\bf{k}}}\right].
\end{eqnarray}
Here, $\hbar {{\bf v}}_{\bf k} = \nabla_{\bf k}{\epsilon}_{\bf k}$ is the band velocity and we have defined $D_{\bf k} = D({\bf B},{\bf \Omega}_{\bf k}) \equiv [1+\frac{e}{\hbar}({\bf B}\cdot{\bf \Omega}_{\bf{k}})]^{-1}$.   
The group velocity of carriers in Eq.~\eqref{eom_r} consists of two BC-dependent terms: the ${\bf E}\times {\bf \Omega}_{\bf k}$ term gives rise to the anomalous Hall effect (AHE) \cite{Xiao2010}, while the $({{\bf v}}_{\bf{k}}\cdot {\bf \Omega}_{\bf k}){\bf B}$ term gives rise to the chiral magnetic effect in presence of a finite chiral chemical potential in WSM \cite{kim_kim}. 
In Eq.~\eqref{eom_k}, the first two terms denote the Lorentz force, whereas the third $({\bf E}\cdot{\bf B}){\bf  \Omega_{\bf{k}}}$ term manifests the effect of the chiral anomaly leading to negative MR \cite{Son_Spivak2013} in WSM.  

The modified EOM also changes the phase-space volume by a factor $D_{\bf k}$, i.e., $[d{\bf k}] \to D_{\bf k} \times [d{\bf k}]$. 
Here, $[d{\bf k}]$ is the shorthand for $d{\bf k}/(2\pi)^{3}$. To counter this changed phase-space volume, so that the number of states in the volume element is preserved, the density of phase-space is multiplied by $\inv{D_{\bf k}}$. This factor needs to be included  
whenever the wave-vector summation is converted in an integral over the BZ \cite{xiao_niu, duval}.

The dynamics of the position and wave-vector-dependent NDF, $g_{\bf r, k}$ is described by the Boltzmann kinetic equation given by \cite{ashcroft}
\begin{equation}\label{bte}
\frac{\partial g_{\bf r, k}}{\partial t} + \dot{\bf{r}}\cdot {\bf \nabla}_{\bf{r}}~ g_{\bf{r},\bf{k}} +\dot{\bf{k}}\cdot{\bf\nabla}_{\bf{k}}~g_{\bf{r},\bf{k}}=I_{\rm coll}\{g_{\bf{r},\bf{k}}\}~,
\end{equation}
where $I_{\rm coll}\{g_{\bf{r},\bf{k}}\}$ is the collision integral. Using the relaxation time approximation for the collision integral in the steady state and homogeneous field ($g_{\bf{r},\bf{k}} \to g_{\bf{k}}$) , the NDF kinetic equation reduces to 
\begin{equation}\label{bte_2}
\dot{\bf{k}}\cdot{\bf\nabla}_{\bf{k}}~g_{\bf{k}} =-\dfrac{g_{\bf{k}}-f_{\rm eq}}{\tau_{\bf{k}}}~,
\end{equation}
where $\tau_{\bf k}$ is the effective intranode relaxation time \cite{ashcroft,PhysRevB.90.165115} and $f_{\rm eq} \equiv f_{\rm eq}({\epsilon}_{\bf k},\mu,T) = (e^{\beta({\epsilon}_{\bf k} - \mu)} + 1)^{-1}$ is the equilibrium Fermi-Dirac distribution function with $\beta^{-1} \equiv k_{B}T$. We consider only the impact of the intranode scattering for tilted nodes. The impact of the internode scattering and its inclusion in isotropic nodes is discussed in Sec. \ref{internode} later in this paper.
Furthermore,  we will consider $\tau_{\bf k}$ to be an isotropic constant ($\tau_{\bf k} \to \tau_\mu$) for a given Fermi energy, in the rest of the paper. 
The energy dependence of $\tau_\mu$ is discussed briefly in Appendix~\ref{secF}.

Substituting 
Eqs.~\eqref{eom_r} and \eqref{eom_k} in Eq.~\eqref{bte_2}, we obtain an approximate NDF, up to first order in ${\bf E}$, to be 
\begin{multline} \label{ndf}
g_{\bf k} = f_{\rm eq} + \bigg[ -e {\bf E}
\cdot \bigg({\bf v}_{\bf k} + \frac{e {\bf B}({\bf v}_{\bf k}\cdot{\bf \Omega}_{\bf k})}{\hbar}\bigg) D_{\bf k}\tau_\mu + {\bf v}_{\bf k} \cdot{\bf \Gamma}\bigg]
\\
\times \bigg(-\dfrac{\partial f_{\rm eq}}{\partial {\epsilon}_{\bf k}}\bigg)~.
\end{multline}
Here, the ${\bf v}_{\bf k} \cdot {\bf \Gamma}$ term accounts for the impact of the ``Lorentz-force" in modifying the NDF \citep{kim_kim, jacoboni}. The Lorentz-force-induced modification in the NDF has terms depending on $\omega_c \tau_\mu$. Recall that the cyclotron frequency is given by $\omega_c = e B/m$, where $m$ denotes the ``inertial" mass of the carriers. In case of Dirac systems, the ``inertial mass" turns out to be density dependent and it is given by $m = \mu/v_F^2$, where $v_F$ is the Fermi velocity of the carriers. Thus, in Dirac systems, we have $\omega_c = e B v_F^2/\mu$. 
In this paper,  we will only focus on the BC-connected conductivity, while not worrying about the Lorentz force contribution to the conductivity. Strictly speaking, for the case of only intranode scattering, the cyclotron motion can be neglected if $(\mu \tau_\mu/\hbar)^2 \ll 1$ \cite{PhysRevB.96.235134}.

\subsection{Magnetoconductivity}
Using the definition of current 
\be\label{current1}
{\bf j}^{e}=-e\int [d{\bf k}]\inv{D_{\bf k}}~\dot{{\bf r}}~g_{\bf k}~,
\ee 
and substituting Eqs.~\eqref{eom_r} and ~\eqref{ndf} in Eq.~\eqref{current1}, yields the following expression for the BC-dependent electrical conductivity matrix:
\begin{multline}\label{elec_cond}
\sigma_{ij}^{\rm total} =-\frac{e^{2}}{\hbar}\int[d{\bf k}]~\epsilon_{ijl}\Omega_{\bf k}^{l}~f_{\rm eq}+e^{2}\tau_\mu
\\
\times \int [d{\bf k}] D_{\bf k}
\bigg({v}_{i}  + \dfrac{eB_{i}}{\hbar}({\bf v}_{\bf k}\cdot {\bf \Omega}_{\bf k})\bigg)
\\
\times \bigg({v}_{j}  + \dfrac{eB_{j}}{\hbar}({\bf v}_{\bf k}\cdot {\bf \Omega}_{\bf k})\bigg)\bigg(-\dfrac{\partial f_{\rm eq}({\epsilon}_{\bf k},\mu)}{\partial {\epsilon}_{\bf k}}\bigg)~.
\end{multline}
Here, ${v}_{i}$ denotes the $i$th component of ${{\bf v}}_{\bf k}$, $\epsilon_{ijl}$ is the Levi-Civita antisymmetric tensor, and $\Omega_{{\bf k}}^{l}$ is simply the $l$th component of the BC.

The conductivity in Eq.~\eqref{elec_cond} can be expressed as $\sigma_{ij}^{\rm total} = \sigma_{ij}^{\rm AHE} + \sigma_{ij}$. Here, 
the first term is the ``intrinsic anomalous" Hall effect \cite{burkov_AHE, axionic_FT,Haldane2004}. These intrinsic anomalous responses arise primarily from a finite BC, and are completely independent of the scattering timescale $\tau_\mu$.

The second term in Eq.~\eqref{elec_cond} is symmetric under the exchange of the indices, $i \to j$ and $j \to i$,, i.e., $\sigma_{ij} = \sigma_{ji}$. 
The BC-dependent velocity correction in one of the last terms in $\sigma_{ij}$ arises from the ${\bf E} \cdot {\bf B}$ terms in Eq.~\eqref{eom_k}, while the other originates from the chiral magnetic term of Eq.~\eqref{eom_r}. 
For parallel electric and magnetic fields in a WSM,  Eq.~\eqref{elec_cond} leads to finite negative MR \cite{Nielsen1983, Son_Spivak2013} quadratic in the magnetic field. This is a relatively well-established transport signature for ideal WSM \cite{Kim2013, Huang2015}. Additionally, this term also leads to the planar Hall effect in WSM \cite{Burkov2017, Nandy2017, Nitesh2018, Yang18}, in which a Hall voltage is generated in the plane of the electric and magnetic fields, as long as they are not parallel or perpendicular to each other.

In this paper, we focus on the MC in tilted type-I and type-II WSM, systematically calculating the anisotropic conductivity matrix including the zeroth order, linear, and quadratic terms. Expanding Eq.~\eqref{elec_cond} in powers of $B$, the conductivity can be expressed as $\sigma_{ij} = \sigma^{(0)}_{ij} + \sigma_{ij}^{(1)} + \sigma^{(2)}_{ij} + \cdots$, where the superscript refers to the order of the magnetic field. Thus, $\sigma^{(0)}$ denotes the Drude contribution, $\sigma^{(1)}$ denotes the linear and $\sigma^{(2)}$ the quadratic contribution in the magnetic field.

The expression for the $B$-linear contribution is explicitly given by $\sigma_{ij}^{(1)} = \int [d{\bf k}] \sigma^{(1)}_{i j {\bf k}}$, where we have defined,  
\be \label{LinB}
\sigma_{ij{\bf k}}^{(1)}= -\frac{e^3 \tau_\mu}{\hbar} 
\left[(B_{i}v_{j}+B_{j}v_{i})({\bf v}\cdot {\bf \Omega})
-({\bf \Omega}\cdot {\bf B})v_{i}v_{j}
\right] \dfrac{\partial f_{\rm eq}}{\partial {\epsilon}_{{\bf k}}}~. 
\ee
The $B$-linear magnetoconductivity in tilted WSM, was earlier demonstrated numerically in Ref.~[\onlinecite{G_Sharma2017a}]. However, a thorough analytical treatment for all possible cases, is still lacking in the literature.
Evidently, from Eq.~\eqref{LinB}, we have $\sigma_{ij}^{(1)}(B) = \sigma_{ji}^{(1)}(B)$, and $\sigma_{ij}^{(1)}(B)$ changes sign on reversing the direction of ${\bf B}$. Such linear terms are, 
in principle, forbidden by the Onsagar-Casimir reciprocity relations \cite{PhysRev.37.405,RevModPhys.17.343,PhysRevB.86.155118}: $\sigma_{ij}(B) = \sigma_{ji}(-B)$, which are valid for systems with TRS. However, in case of a tilted Weyl node, the TRS is broken by the tilt, allowing for the possibility of finite conductivity contributions from such $B$-linear terms.  Note that the tilt of the WNs breaks its TRS even if it is located at the origin ($Q =0$).

Similarly, the quadratic-$B$ contribution is explicitly given by 
\begin{multline} \label{Bquad}
\sigma_{ij}^{(2)}= \frac{e^4 \tau_\mu}{\hbar^2}
\int [d{\bf k}]
\Big[v_{i} v_{j}\left({\bf B}\cdot {\bf \Omega}\right)^{2}
-({\bf B}\cdot {\bf \Omega})
(v_{j}B_{i}+v_{i}B_{j})
\\
\times({\bf v}\cdot {\bf \Omega})
+B_{i}B_{j}({\bf v}\cdot{\bf \Omega})^{2}\Big]
\bigg(-\dfrac{\partial f_{\rm eq}({\epsilon}_{{\bf k}},\mu)}{\partial {\epsilon}_{{\bf k}}}\bigg)~.
\end{multline}
Evidently, $\sigma_{ij}^{(2)} (B) = \sigma_{ji}^{(2)} (B)$.

In order to obtain analytical expressions for the MC matrix, we base our calculations on the simple low-energy effective Hamiltonian with a pair of tilted WNs, separated in the momentum space. Additionally, for the ease of calculation, we resort to zero-temperature, where the derivative of the Fermi function reduces to a Dirac-delta function. The finite-temperature results can be easily obtained by integrating the zero-temperature result \cite{ashcroft},
\begin{equation}
\sigma(\mu,T) = -\int_{-\infty}^{\infty} \sigma(E,T=0)~ \frac{\partial f_{\rm eq}(E,\mu)}{\partial E}~dE.
\end{equation}

In most of the experimentally discovered tilted WSM, the tilt axis (which we denote by ${\hat{\bf R}}$) is parallel to ${\bf Q}$, and this is what we will work with in the rest of the paper. We consider the cases of ${\bf B} \parallel \hat{\bf R}$, and ${\bf B} \perp \hat{\bf R}$ separately, and explicitly calculate the full conductivity matrix for the three cases of (1) isotropic WSM, (2) tilted type-I WSM, and (3) ``tilted over" type-II WSM.

\section{Isotropic Weyl nodes}
\label{Iso}
The effective low-energy continuum Hamiltonian for TRS broken WSMs with a pair of nodes separated along the $z$ axis, is given by
\be\label{Ham_iso}
\mathcal{H}_{s}({\bf k}) = s\hbar v_{F} ~{\bf \sigma}\cdot({\bf k}-s Q \hat{\bf e}_{z})~.
\ee
Here $s = \pm 1$ is the chirality eigenvalue for a given WN, $Q$ is the distance of the WN from origin and $\sigma = \{\sigma_x,\sigma_y,\sigma_z \}$ is the vector comprising the three Pauli matrices. 
The Hamiltonian in Eq.~\eqref{Ham_iso}, preserves the particle-hole symmetry and is Lorentz invariant. 
The BC corresponding to Eq.~\eqref{Ham_iso} can be analytically calculated and it is given by \cite{PhysRevB.95.165135}
\be \label{BerryCurv}
{\bf \Omega}_{s}({\bf k})=\mp s \frac{{\bf k}-s {\bf Q}}{2~|{\bf k}-s {\bf Q}|^{3}}~,
\ee
where the $-$ ($+$) sign denotes the conduction (valence) band. Interestingly, the BC is impervious to anisotropy in the system, and Eq.~\eqref{BerryCurv} holds for tilted  type-I and ``tilted over" type-II WSM as well.

In the absence of a magnetic field, the conductivity matrix $\sigma^{(0)}_{ij}$ for each of the isotropic WNs, turns out to be diagonal.  
Moreover, the longitudinal conductivities in the absence of magnetic field (or Drude conductivity, $\sigma^{(0)}_{ii} \equiv \sigma_{\rm D}$) is identical in all directions. For each node, it is given by
\be \label{drude_iso}
\sigma_{\rm D}(\mu)=\frac{4 \pi}{3}\dfrac{e^{2}}{h} \dfrac{\mu^{2} \tau_\mu}{h^{2} v_{F}} ~.
\ee
The Drude conductivity varies as a square of the chemical potential [barring the $\mu$ dependence of $\tau_\mu(\mu)$], vanishing on approach to the Weyl point.  This is expected as the density of states in an isotropic WSM also varies as $\mu^2$ and vanishes on approaching the Weyl point. Equation \eqref{drude_iso} has also been derived earlier using the Kubo formula \cite{PhysRevB.93.085426}.

\subsection{Magnetic field perpendicular to the tilt axis (${\bf B} \perp \hat{\bf R}$)}
Now let us switch on the magnetic field in the $x$-$y$ plane, perpendicular to the node separation, at an angle $\phi$ measured in an anti-clockwise sense from the $x$ axis. In this configuration, $\sigma^{(1)}$ is identically zero. The lowest-order $B$ correction in the conductivity is quadratic. 
The in-plane longitudinal component of the MC for each node is obtained to be
\be\label{sigma_xx_2_iso}
\sigma_{xx}^{(2)} (\phi)=8\sigma_{0}\cos^{2}\phi+\sigma_{0}\sin^{2}\phi~,
\ee
and as expected, $\sigma_{yy}^{(2)} ( \phi, \mu) = \sigma_{xx}^{(2)} (\pi/2-\phi, \mu)$. Here 
we have defined 
\be \label{sigma0}
\sigma_{0}=\dfrac{e^{2}\tau_\mu}{8\pi^2}\dfrac{\hbar v_{F}^{3}}{15 \mu^{2}}\left(\dfrac{eB}{\hbar}\right)^{2},
\ee
which is independent of the chirality of the WNs. This $B^2$ dependence of the MC has been experimentally observed in the MR of Cd$_3$As$_2$ \cite{Li2016}, topological insulators \cite{Kim2013}, and Na$_3$Bi \cite{Xiong2015}, among others. A similar $\mu^{-2}$ dependence of the longitudinal conductivity has been reported to arise due to chiral anomaly in  Ref.~[\onlinecite{Son_Spivak2013}]. The origin of this can be traced back to the $1/k^2$ dependence of the BC in WSM. 

The other diagonal component, $\sigma_{zz}^{(2)} = \sigma_{0}$. Among the off-diagonal components, $\sigma^{(2)}_{xz} = \sigma^{(2)}_{yz}=0$, 
and 
\be \label{sigma_xy_2_iso}
\sigma_{yx}^{(2)}(\phi)=7\sigma_{0}\sin\phi \cos\phi~.
\ee
This is the planar Hall effect in WSM: generation of a Hall voltage in the $B$-$E$ plane, perpendicular to the applied $E$ direction, studied in detail in Ref.~[\onlinecite{Nandy2017}]. 

The quadratic $B$ dependence of the longitudinal and the planar Hall conductivity can also be expressed as \cite{Nandy2017}, 
\bea \label{Eq20}
\sigma_{xx}&=&\sigma_{\perp}+\Delta\sigma\cos^2\phi~,\\
\sigma_{xy}&=&\Delta\sigma\sin\phi\cos\phi~. \label{Eq21}
\eea
Here, $\Delta\sigma=\sigma_{\parallel}-\sigma_{\perp}$ along with $\sigma_{\parallel}\equiv \sigma_{xx}(\phi=0)=\sigma_{\rm D}+8\sigma_{0}$ and $\sigma_{\perp} \equiv \sigma_{xx}(\phi=\pi/2)=\sigma_{\rm D}+\sigma_{0}$. 
We emphasize that we find explicit quadratic-$B$ corrections in $\sigma_{\perp}$ via the $\sigma_0$ term,  which has been missed in a few earlier theoretical works \cite{Nandy2017}. In terms of resistivity, this translates to $\rho_{\perp}=\rho_{\rm D}-{\sigma_{0}}/{\sigma^2_{\rm D}}$, where the second term originates from the BC-dependent phase-space factor.  
It is likely to be useful in correctly interpreting experimental results related to planar Hall effect in WSM \cite{Nitesh2018}. For example, in Ref.~[\onlinecite{Nitesh2018}], an extra magnetic-field-dependent term has been added by ``hand"  in the $\rho_{\perp}$ 
to correctly interpret the planar Hall results.

\subsection{Magnetic field along the tilt axis (${\bf B} \parallel \hat{\bf R}$)}
Here, we consider the case when the magnetic field is applied in the direction of the tilt axis which also coincides with the node-separation, i.e., along the $z$ direction. In this case also all the $B$-linear components of the conductivity matrix are identically zero. The quadratic-$B$ correction terms are given by $\sigma_{zz}^{(2)} = 8\sigma_{0}$, and $\sigma_{xx}^{(2)}= \sigma_{yy}^{(2)}= \sigma_{0}$. All other off-diagonal components of the conductivity matrix are zero. 

Note that in this section all the results are for a single WN. For the case of multiple nodes, these single-node expressions have to be multiplied by the total number of nodes in the WSM. Following, we will explicitly consider the case of WSM, with a pair of WNs.

\begin{figure}[h!]
\includegraphics[width =\linewidth]{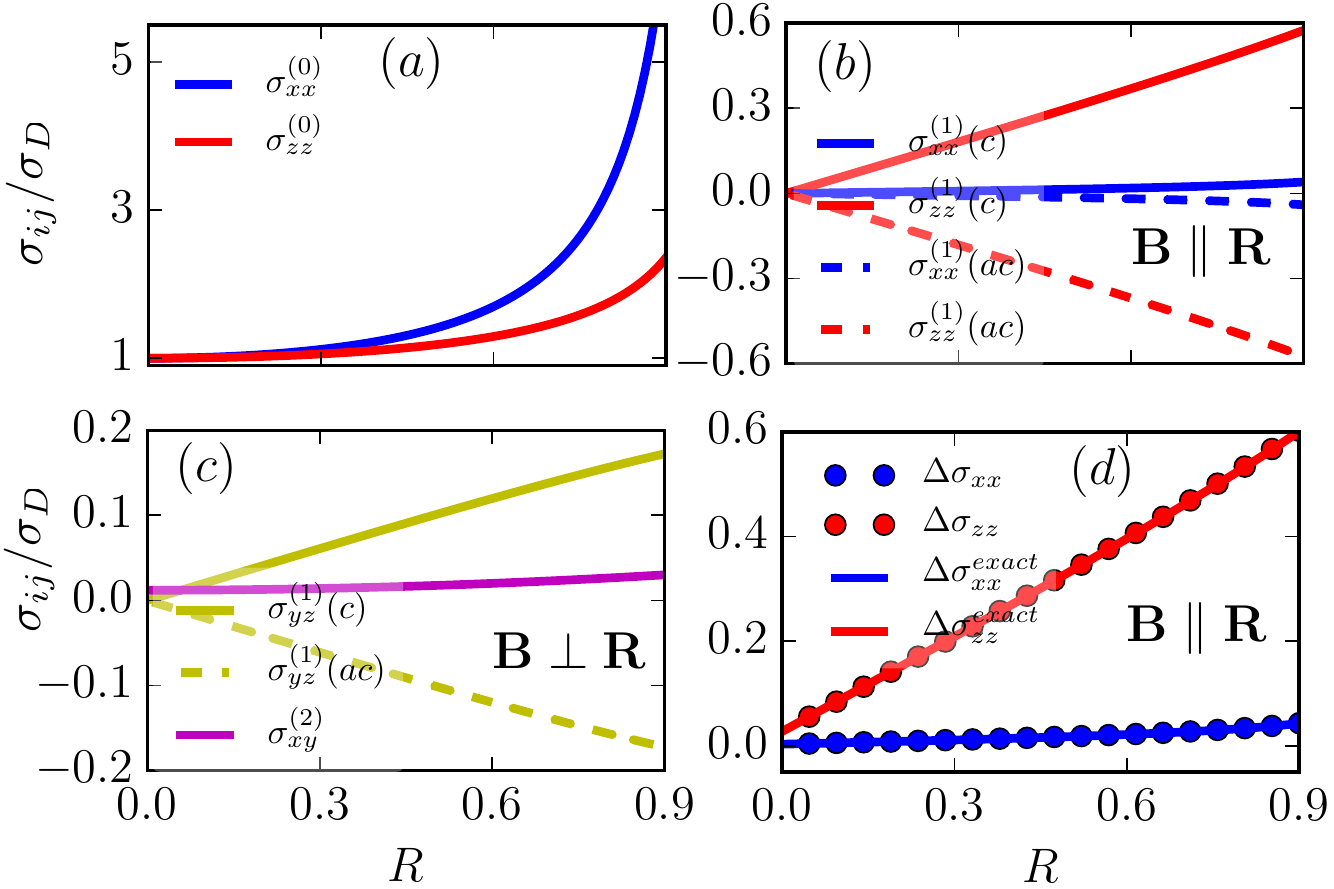}
\caption{The tilt dependence of the magneto-conductivity for a type-I WSM with a pair of oppositely tilted Weyl nodes ($R_+= -R_-= R$). (a) The variation of the Drude component $\sigma^{(0)}_{ii}$ with $R$.  (b) The variation of the $B$-linear diagonal term in the magneto-conductivity with $R$ for the case of ${\bf B} \parallel \hat{\bf R}$. Interestingly, the linear correction is sensitive to the sign of $R$, and behaves differently for the cases when the tilt is clockwise (labelled as $c$ for the case of $s =1, R_+<0$) as opposed to anticlockwise (labelled as $ac$ for the case of $s =1, R_+>0$). (c) The linear and quadratic correction to the transverse conductivities for ${\bf B} \perp \hat{\bf R}$.
(d) The correction to magneto-conductivity including $B$-linear and quadratic-$B$ terms ($\Delta\sigma = \sigma^{(1)} + \sigma^{(2)}$, denoted by dotted line), compared with the numerically exact calculation (to all orders in $B$, denoted by solid lines) based on Eq.~\eqref{elec_cond}. 
We have used the following parameters: $\mu$ = 0.1 eV, $v_{F}=10^{6}$ m/s, $\tau_\mu= 10^{-12}$ sec, $B = 4$ T and  $\phi = \pi/4$.}
\label{fig:fig_2}
\end{figure}

\section{Tilted Type-I Weyl Semimetal}
\label{type-I}
Having described the BC-induced MC in isotropic WSM, we turn our focus on tilted type-I WSM. 
Adding a tilt along the $z$ axis in Eq. \eqref{Ham_iso}, the anisotropic low-energy continuum WSM Hamiltonian is given by \cite{Zyuzin_A_A2016}
\be\label{Ham_type_I}
\mathcal{H}_{s}({\bf k}) = \hbar C_{s}(k_{z}-s Q)+s\hbar v_{F} {\bf \sigma}\cdot({\bf k}-s Q \hat{\bf e}_{z})~.
\ee
Here, $C_{s}$ is the tilt velocity, which can in principle be different for different nodes. 
Following Ref.~[\onlinecite{Zyuzin_A_A2016}], we allow for the tilt of the two nodes to be different.   
However, we should bear in mind that most of the experimentally observed WSM are oppositely tilted ($C_+ = -C_-$). 
This Hamiltonian corresponds to type-I WSM only for small tilt velocity, $|C_{s}|<v_{F}$, such that the Fermi surface encloses only a electron ($\mu >0$) or a hole ($\mu<0$) pocket. A schematic of a tilted type-I WSM node with $\mu > 0$, is shown in Fig.~\ref{f1}(a). It is useful to define the tilt parameter as $R_s\equiv C_{s}/v_{F}$ and for type-I WSMs, $|R_s|<1$.

On account of the anisotropic dispersion, the Drude conductivity is different in different direction. The diagonal component of the Drude conductivity 
perpendicular to the node separation is
\be\label{sigma_xx_0_I_tilt}
\sigma_{xx}^{(0)}=\sum_s \dfrac{3\sigma_{\rm D}}{4R_s^{3}}\left[\dfrac{2R_s}{(1-R_s^{2})} + \ln\left(\dfrac{1-R_s}{1+R_s}\right)\right]~,
\ee
and, $\sigma_{yy}^{(0)} = \sigma_{xx}^{(0)}$. 
It is easy to check that Eq.~\eqref{sigma_xx_0_I_tilt} reduces to Eq.~\eqref{drude_iso} in the $R_s \to 0$ limit. As an additional check, 
we note that it is also identical to Eq.~(23) of Ref.~[\onlinecite{carbotte}]. As expected, $\sigma_{xx}^{(0)} \propto \mu^2$, as in the isotropic case. 
The Drude conductivity along the direction of the tilt is given by 
\be \label{sigma_zz_0_I_tilt}
\sigma_{zz}^{(0)}=\sum_s\dfrac{3\sigma_{\rm D}}{2R_s^{3}}\left[-2R_s -\ln\left(\dfrac{1-R_s}{1+R_s}\right) \right]~.
\ee
Since both Eqs.~\eqref{sigma_xx_0_I_tilt} and \eqref{sigma_zz_0_I_tilt} are even functions of $R_s$, the total Drude contribution for the case with different tilt directions just adds up. All other off-diagonal elements of the Drude conductivity are zero. Figure~\ref{fig:fig_2}(a) shows the $R_s$ dependence of $\sigma^{(0)}_{xx}$ and $\sigma^{(0)}_{zz}$. Evidently, $\sigma^{(0)}_{xx}$ is more sensitive to $R_s$ as compared to $\sigma^{(0)}_{zz}$.

To calculate the conductivity in presence of a magnetic field, we consider two different cases: $B$ applied  along the direction of the tilt axis and perpendicular to it, in the subsections following.

\subsection{Magnetic field along the tilt axis (${\bf B} \parallel \hat{\bf R}$)}
Unlike the isotropic case, a tilted WN gives rise to a $B$-linear contribution, in addition to the quadratic-$B$ contribution in the conductivity matrix. The $B$-linear conductivity component along the tilt direction is found to be
\be \label{sigma_zz_I_1_par}
\sigma_{zz}^{(1)}= \sum_s \dfrac{s \sigma_{1}}{3R_s^{4}}\left[2R_s\left(3-5R_s^{2}-3R_s^{4}\right) + 3(R_s^{2}-1)^{2} \delta_s\right].
\ee
Here we have defined,
\be 
\sigma_{1}=\frac{e^{2}\tau_\mu}{(2\pi)^{3}}\frac{\pi v_{F}}{\hbar} \frac{eB}{\hbar}, 
\ee
and 
\be
\delta_s = \ln \left( \dfrac{1-R_s}{1+R_s}\right)~.
\ee 
The linear conductivity in the plane perpendicular to the tilt axis is given by,  $\sigma_{xx}^{(1)} = \sigma_{yy}^{(1)}$, 
where 
\be \label{sigma_xx_I_1_par}
\sigma_{xx}^{(1)}=\sum_s\dfrac{ s \sigma_{1}}{6R_s^{4}}\bigg[4R_s^{3}-6R_s
-3(1-R_s^{2})\delta_s\bigg]~.
\ee
There are three very interesting things about this $B$-linear MC term. 
(1) It is independent of the Fermi energy barring the $\mu$ dependence of $\tau_\mu$. (2) The linear MC depends on the relative orientation of the tilt of the different nodes. 
As evident from Eq.~\eqref{sigma_zz_I_1_par}, it vanishes in the case of opposite chirality nodes tilted in the parallel direction, i.e., for $R_+ = R_-$, the sum over $s \to 0$. 
However, if the nodes are oppositely tilted ($R_+ = -R_-$), as in most of the experimentally observed WSM, the different chirality terms just add up. 
3) The sign of the linear term is dictated by the relative orientation of the chirality and the tilt axis, $\sigma_{xx}^{(1)} \propto s \times {\rm sign}(R_s)$. For example, in the case of a WSM with a pair of oppositely tilted WN, if the tilt is in the 
anti-clockwise (clockwise) direction to the $z$ axis for the positive chirality ($s=+1$) node, then the overall sign of the $B$-linear MC is negative (positive), as shown in Fig.~\ref{fig:fig_2}(b). For these $B$-linear corrections in the longitudinal terms, the corresponding magneto-current turns out to be 
${\bf j} \propto (\hat{\bf R}\cdot{\bf B}) \hat{\bf E}$.

These $B$-linear MC correction in tilted WSM,  were first reported in tilted WSM numerically, based on a lattice model of WSM in  
Ref.~[\onlinecite{G_Sharma2017a}]. More recently, $B$-linear terms were also explored in Ref.~[\onlinecite{Zyuzin_V_A2017}] for small $R_s$ values, and shown to arise from 
the TRS breaking tilt of the Weyl nodes and chiral anomaly. In an earlier work \cite{Cortijo2016}, similar $B$-linear corrections were shown to arise in TRS broken WSM, 
where the TRS breaking was due to the cubic band-bending terms ($ \propto  k_z^3 \sigma_z$) in the Hamiltonian. 

The quadratic-$B$ contribution to the conductivity is given by
\be\label{sigma_zz_I_2_par}
\sigma_{zz}^{(2)}=\sum_s 8\sigma_{0}~,
\ee
where $\sigma_0$ is defined in Eq.~\eqref{sigma0}. This turns out to be independent of the tilt and is identical to the isotropic case. 
Similarly, the other two diagonal terms for quadratic-$B$ contribution are given by 
\be \label{sigma_xx_I_2_par}
\sigma_{xx}^{(2)} = \sigma_{yy}^{(2)}= \sum_s\sigma_{0}.
\ee
All the off-diagonal terms in the conductivity matrix are identically zero in this case, as in the isotropic case.

\subsection{Magnetic field perpendicular to the tilt axis (${\bf B} \perp \hat{\bf R}$)}
Unlike the isotropic case, for $B$ applied in the $x$-$y$ plane (${\bf B} \perp \hat{\bf R}$) in a WSM with tilt, 
there are $B$-linear corrections to the off-diagonal components to the conductivity. 
The finite off-diagonal components of the conductivity are given by 
\be
\sigma_{xz}^{(1)}=\sigma_{2} \cos\phi~~~{\rm and}~~~~\sigma_{yz}^{(1)}=\sigma_{2}\sin\phi~,
\ee
where we have defined $\sigma_{2}$ as 
\be\label{sigma_2_I_per}
\sigma_{\rm 2}=\sum_s\dfrac{s\sigma_{1}}{6R_s^{4}}\left[-2R_s\left(3-5R_s^{2}+6R_s^{4}\right) - 3\left(1-R_s^{2}\right)^2 \delta_s \right]~.
\ee
All the other $B$-linear terms are zeros, i.e., $\sigma_{xx}^{(1)} = \sigma_{yy}^{(1)}= \sigma_{zz}^{(1)}= \sigma_{xy}^{(1)}= 0$. Here also, 
the $B$-linear Hall conductivities are independent of the chemical potential (barring the $\mu$ dependence of $\tau_\mu$). Since Eq.~\eqref{sigma_2_I_per} is an odd function in both $R_s$ and $s$, 
the $B$-linear Hall conductivities vanish if the two nodes are tilted parallel to each other. Additionally, the sign of the these terms determines the 
tilt direction of the positive chirality node [see Fig.~\ref{fig:fig_2}(c)]. These $B$-linear components of the transverse conductivities correspond to the magneto-current of the form  ${\bf j} \propto ({\bf E}\cdot\hat{\bf R}){\bf B}$ (for $\sigma_{xz}^{(1)}$ and $\sigma_{yz}^{(1)}$) or alternately ${\bf j} \propto ({\bf E}\cdot{\bf B}) \hat{\bf R}$ (for $\sigma_{zx}^{(1)}$ and $\sigma_{zy}^{(1)}$).

In addition to generating the $B$-linear terms above, the tilt in WSM also modifies the quadratic-$B$ terms in the conductivity.   
The diagonal terms are given by $\sigma_{yy}^{(2)}(\phi) = \sigma_{xx}^{(2)}(\pi/2-\phi)$, where 
\be\label{sigma_xx_I_2_per}
\sigma_{xx}^{(2)}(\phi)=\sigma_{0} \sum_s \left[(8+13R_s^{2})\cos^{2}\phi+ \sin^{2}\phi\right]~,
\ee
and 
\be\label{sigma_zz_I_2_per}
\sigma_{zz}^{(2)}= \sigma_{0} \sum_s (1+7R_s^{2})~.
\ee
Note that this is similar to the isotropic case, with some modifications induced by the tilt parameter. 
The off-diagonal term of the quadratic-$B$ conductivity in the $x$-$y$ plane is given by 
\be \label{sigma_xy(2)}
\sigma_{xy}^{(2)}= \sigma_0 \sin\phi \cos\phi \sum_s (7+13R_s^{2})~.
\ee
All the other off-diagonal components are zero, $\sigma^{(2)}_{xz} = \sigma^{(2)}_{yz} = 0$. The angular dependence of the longitudinal conductivity ($\sigma_{xx}^{(2)}$)
and the planar Hall conductivity ($\sigma_{xy}^{(2)}$) for type-I WSM are shown in Figs \ref{fig:fig_5}(a) and \ref{fig:fig_5}(c), respectively.

\begin{figure}[t]
\includegraphics[width = \linewidth]{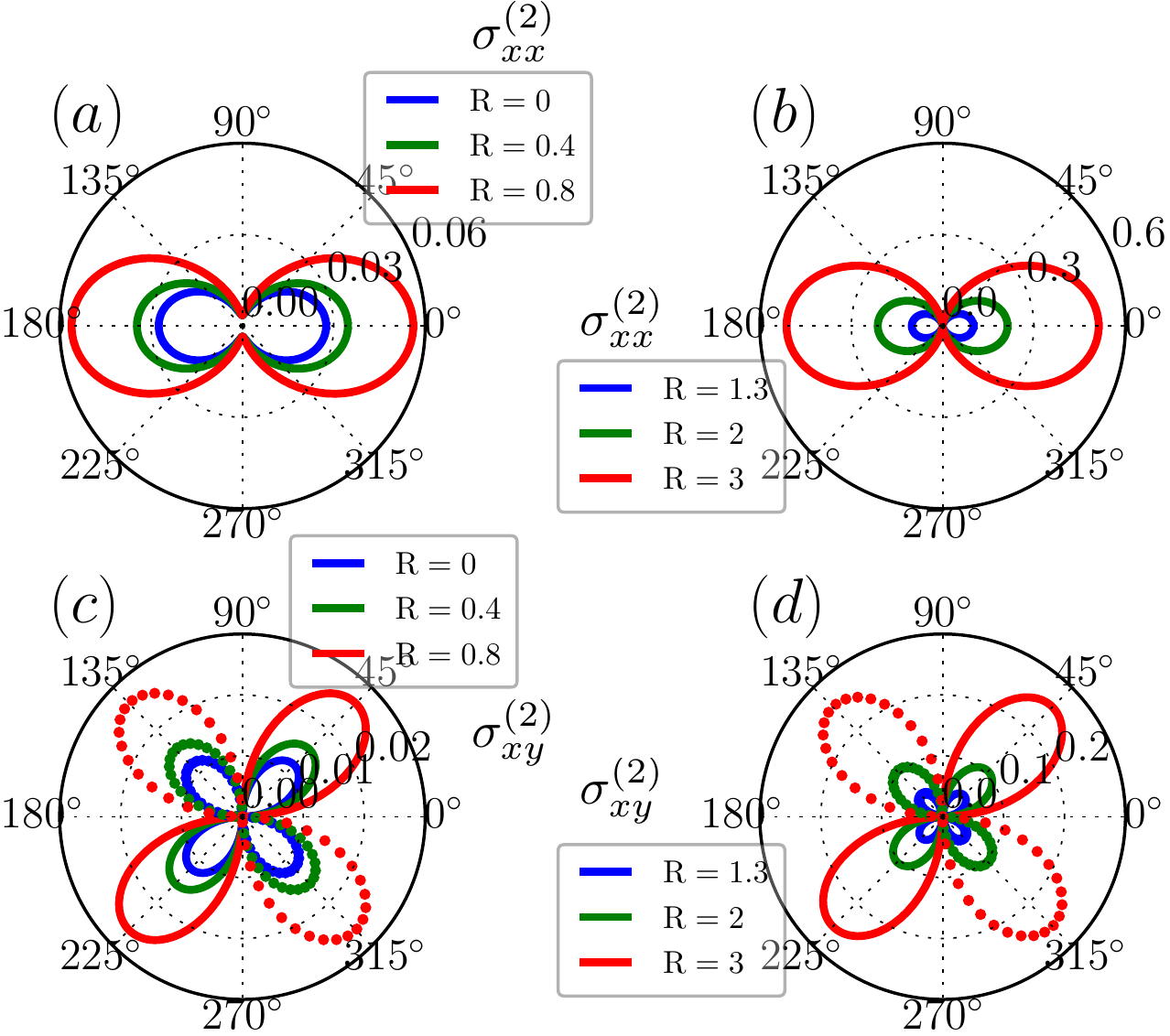}
\caption{(a), (b) The quadratic-$B$ correction to MC with finite field angle $\phi$, for a type-I and a type-II WSM, respectively, for different values of the tilt angle. (c), (d) show the $\sin(2 \phi)$ dependence of the planar Hall conductivity for a type-I, and a type-II WSM, respectively. All the conductivities here are in units of $\sigma_{\rm D}$, for ease of comparison. The dotted curves represent the negative values of the corresponding quantities. All the parameters used are identical to that of Fig.~\ref{fig:fig_2} (for type-I) and Fig.~\ref{fig:fig_3} (for type-II).
\label{fig:fig_5} }
\end{figure}

\section{type-II WSM}
\label{type-II}
Having discussed the MC of type-I WSM, we now proceed to discuss the MC of type-II WSM. In type-II WSM, we have $|R_s| >1$ or $|C_s| > v_F$, leading to over tilted WNs, and the presence of unbounded electron and hole pockets at the Fermi energy. Consequently, in type-II WSM, there is a finite contribution to all physical properties arising from both the conduction and valence bands, in stark contrast to a type-I WSM.

In type-II WSM, the unbounded nature of the Fermi surface poses a challenge for the calculation of any physical quantity. However, in condensed matter systems, the real band structure of any material is always bounded, and it mimics the ideal WSM band structure obtained from Eq.~\eqref{Ham_type_I} only in a small momentum and energy region in vicinity of the Weyl node. Thus it is only prudent to introduce a seemingly artificial ultra-violet energy (or momentum) cutoff for calculating the physical properties of WSM. 
The cutoff can be chosen in two ways: (1) for Weyl nodes separated in the 
$z$ direction, we can use a cylindrical geometry with the momentum cutoff introduced only along the $z$ direction \cite{Zyuzin_A_A2016}; (2) use spherical symmetry with the ultra-violet cut-off along the radial momentum direction \cite{carbotte}. In this section we focus on the first choice based on cylindrical geometry, with the wave-vector cutoff 
($\Lambda_z$) appearing only in the $z$ direction. However, as an independent check, and for the sake of completeness, we also calculate the MC using  spherical geometry with a radial cutoff in Appendix \ref{appendixA}. See Appendices \ref{appendixB} and \ref{appendixC} for the corresponding details of calculations. Also, note that for type-I WSM, in the absence of a cutoff, both the schemes give identical results as expected.

As usual we start with the zero field component of the conductivity matrix. The longitudinal components of the conductivity matrix are given by ${\sigma}_{xx}^{(0)} = {\sigma}_{yy}^{(0)}$, where 
\begin{multline} \label{sigma_xx_0_sep_II_c}
\dfrac{\sigma_{xx}^{(0)}}{\sigma_{\rm D}} = \dfrac{3}{4}\sum_{s} \bigg[\dfrac{1}{|R_{s}|}
\bigg(\dfrac{R_{s}^2+1}{R_{s}^2-1}-\dfrac{\delta^{1}_{s}}{R_{s}^2}\bigg)
+2s\tilde{Q}\dfrac{R_s^2+1}{R_s^2}{\rm sgn}(R_s)
\\
+\dfrac{R_{s}^{2}-1}{|R_{s}|}\tilde{\Lambda}_k^{2}
-\dfrac{\delta^{2}_{s}}{|R_{s}|^{3}}\bigg]~.
\end{multline}
Here, $\tilde{Q}\equiv Q/k_{F}$, $\tilde{\Lambda}_k\equiv \Lambda_k/k_{F}$ and we have defined the following:
\bea\nn
\delta^{1}_{s}&=&\ln(R_{s}^2 - 1)~,\\\nn
\delta^{2}_{s}&=&\ln\left[R_{s}^2\tilde{\Lambda}_k^{2}-1-2s\tilde{Q}R_{s}\right].
\eea
For the sake of simplicity, in Eq.~\eqref{sigma_xx_0_sep_II_c}  we have neglected terms containing $Q^2/\Lambda_z^2$. The exact result, retaining all orders in $Q/\Lambda_z$, is presented in Appendix \ref{Appendix-type-II}.

In Eq.~\eqref{sigma_xx_0_sep_II_c}, the first term varies quadratically with the chemical potential (barring the $\mu$ dependence of $\tau_\mu$), similar to the case of type-I WSM. The second term varies linearly with $\mu$ and it arises from a finite value of the node separation. The third term, being completely independent of $\mu$,  arises from the ``infinite sea" of electrons in the valence band, and it varies quadratically with the cutoff, similar to the density of states in a type-II WSM. The fourth term arises from the unbounded nature of the Fermi surface, and it involves logarithmic terms in the cutoff.

For a WSM with a pair of nodes having parallel tilt direction ($R_+ = R_-$), the second term in Eq.~\eqref{sigma_xx_0_sep_II_c} which is linear in $\mu$ cancels out and the total Drude conductivity becomes almost independent of $Q$.   
The other diagonal component of the Drude conductivity is given by
\begin{multline}
\dfrac{\sigma_{zz}^{(0)}}{\sigma_{\rm D}}= \frac{3}{2}\sum_{s} \bigg[\dfrac{1}{|R_{s}|}
\bigg(1+R_{s}^2+\dfrac{\delta^{1}_{s}}{R_{s}^{2}}\bigg)
-~2s\tilde{Q}~\dfrac{R_{s}^{4}-1}{R_{s}^{2}}{\rm sgn}(R_{s})
\\
+~\dfrac{(R_{s}^{2}-1)^2}{|R_{s}|}\tilde{\Lambda}^{2}_k
+\dfrac{\delta^{2}_{s}}{|R_{s}|^{3}} \bigg]~.
\end{multline}
All the other off-diagonal components of the zero-field Drude conductivity are identically zero, similar to the case of type-I WSM.

Both the components of the Drude conductivity in a type-II WSM with a pair of oppositely tilted Weyl nodes are shown in Fig.~\ref{fig:fig_4}(a). Unlike the case of type-I WSM [see Fig.~\ref{fig:fig_2}(a)], for a type-II WSM we find that $\sigma_{zz}^{(0)}$ is larger and more sensitive to $R_s$ as compared to $\sigma_{xx}^{(0)}$. 
Having discussed the zero field conductivity of type-II WSM, we now proceed to discuss the $B$ dependent conductivity, considering the two cases: 
(1) ${\bf B} \parallel \hat{\bf R}$ and  (2) ${\bf B} \perp \hat{\bf R}$~.

\subsection{Magnetic field along the tilt axis (${\bf B} \parallel \hat{\bf R}$)}
In this scenario, the diagonal components of conductivity matrix have a finite $B$-linear term. For simplicity, in this section we retain terms only up to first order in $1/\tilde{\Lambda}_k = k_F/\Lambda_z$. The exact results are presented in Appendix \ref{Appendix-type-II}. The $B$-linear term of the diagonal conductivity in the $z$ direction is given by 
\be \label{sigma_zz_II_1_par}
\sigma_{zz}^{(1)} = \sum_{s}\dfrac{s\sigma_{1}}{3R_{s}^{4}}
~{\rm sgn}(R_{s})\left[\mathcal{A}_{z}
-3\mathcal{F}\left(\delta^{1}_{s}+\delta^{2}_{s}\right)
\right],
\ee
where we have defined the following polynomials of $R_{s} $:
\begin{eqnarray}\nn
\mathcal{A}_{z}&=&\left(11- 15R_{s}^2-6R_{s}^4\right)~,\\ \nn
\mathcal{F}&=&(R_{s}^2-1)^2~. 
\end{eqnarray}
The other components of the $B$-linear term in the  diagonal conductivities are given by $\sigma_{xx}^{(1)}=\sigma_{yy}^{(1)}$, where
\be \label{sigma_xx_II_1_par}
\sigma_{xx}^{(1)} = \sum_{s}\dfrac{s\sigma_{1}}{6R_{s}^{4}}
~{\rm sgn} (R_{s})\left(\mathcal{A}_{x}
-3\mathcal{F}^{1/2}\left(\delta^{1}_{s}+\delta^{2}_{s}\right)
\right).
\ee
\begin{figure}[t]
\includegraphics[width = \linewidth]{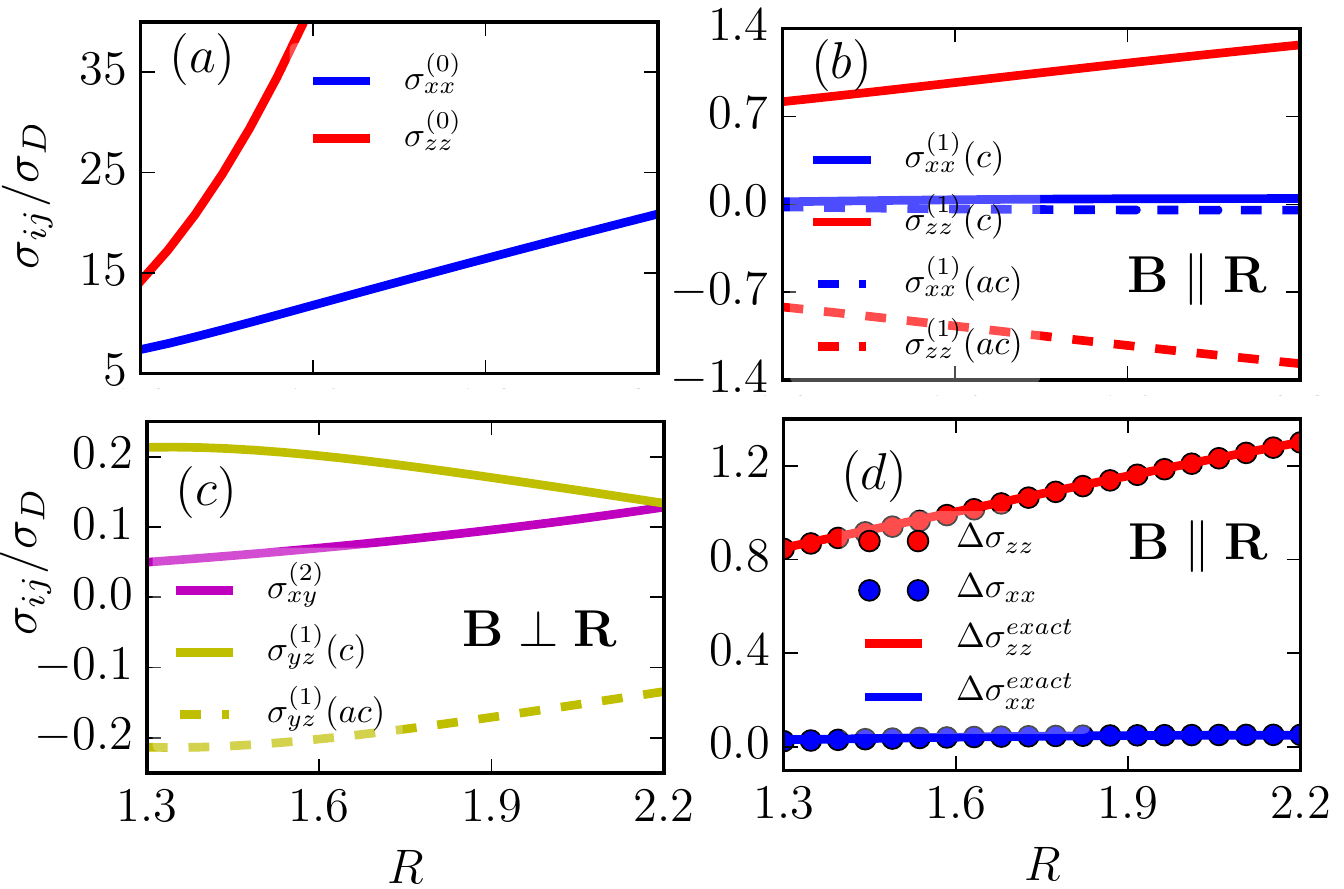}
\caption{The tilt dependence of the magneto-conductivity (in units of $\sigma_{\rm D}$), for type-II WSM, with a pair of oppositely tilted Weyl nodes ($R_+ = -R_-=R$). 
(a) The variation of Drude conductivities $\sigma^{(0)}$ with $R$. Unlike the case of type-I WSM, $\sigma_{zz}^{(0)}$ increases more rapidly then $\sigma_{xx}^{(0)}$. (b) The $B$-linear conductivity ($\sigma_{ij}^{(1)}$) for the case of ${\bf B} \parallel \hat{\bf R}$, as a function of $R$. Solid lines are for a clockwise tilt ($R_+ < 0$, denoted by $c$) relative to the $z$-axis, and dashed lines are for an anti-clockwise tilt ($R_+ > 0$, denoted by $ac$). Panel (c) shows linear and quadratic correction to transverse conductivities when ${\bf B} \perp {\bf R}$ as a function of $R$. In general the linear term is sensitive to the tilt direction. 
(d) The analytically calculated correction to magneto-conductivity upto second order in $B$, $\Delta\sigma = \sigma^{(1)} + \sigma^{(2)}$ (denoted by dotted line), compared with the exact numerical calculation (to all orders in $B$, denoted by solid line) based on Eq.~\eqref{elec_cond}. 
Here we have used the following parameters: $\tilde{Q}=0.1$, $\tilde{\Lambda}_k=4$, $B=4$ T. All other parameters are identical to those in Fig.~\ref{fig:fig_2}.}
\label{fig:fig_4}
\end{figure}

Here, we have defined  
\begin{eqnarray}\nn  
\mathcal{A}_{x} &=& \left(-11 +9R_{s}^2\right)~.
\end{eqnarray} 
There are no finite off-diagonal $B$-linear terms in the conductivity, i.e., $\sigma_{xy}^{(1)}=\sigma_{yz}^{(1)}=\sigma_{zx}^{(1)}=0$. Interestingly, the sign of the linear conductivity terms depends on the orientation of the tilt, relative to the tilt axis, as shown in Fig.~\ref{fig:fig_3}(b). This will show up as anisotropic line-shape in MR experiments for $B <0$, and $B>0$, respectively. Similar to the case of type-I WSM, these magneto-currents corresponding to the $B$-linear conductivity terms, can be expressed as ${\bf j} \propto (\hat{\bf R}\cdot{\bf B}) \hat{\bf E}$.

We now turn our focus on the quadratic-$B$ dependence. Unlike the quadratic-$B$ correction of MC in type-I WSM for ${\bf B} \parallel {\bf R}$, the corresponding terms for type-II WSM are tilt dependent. Keeping terms only up to the first order in $1/\tilde{\Lambda}_k$, the quadratic-$B$ correction to the conductivity along the tilt direction is given by
\be \label{sigma_zz_II_2_par}
\sigma_{zz}^{(2)} = \sum_{s}\dfrac{\sigma_{0}}{2|R_{s}|^5}\left(1-5R_{s}^2+15R_{s}^4+5R_{s}^6\right).
\ee
The quadratic-$B$ correction for the other two diagonal terms is given by $\sigma_{xx}^{(2)}=\sigma_{yy}^{(2)}$, where
\be \label{sigma_xx_II_2_par}
\sigma_{xx}^{(2)} = \sum_{s}\dfrac{\sigma_{0}}{8|R_{s}|^5}\left(-2+5R_{s}^2+5R_{s}^6\right).
\ee
Similar to the case of type-I WSM, all the other off-diagonal components of the conductivity have no quadratic-$B$ terms, i.e., $\sigma_{xy}^{(2)}=\sigma_{yz}^{(2)}=\sigma_{zx}^{(2)}=0$.

\begin{figure}[t]
\includegraphics[width = \linewidth]{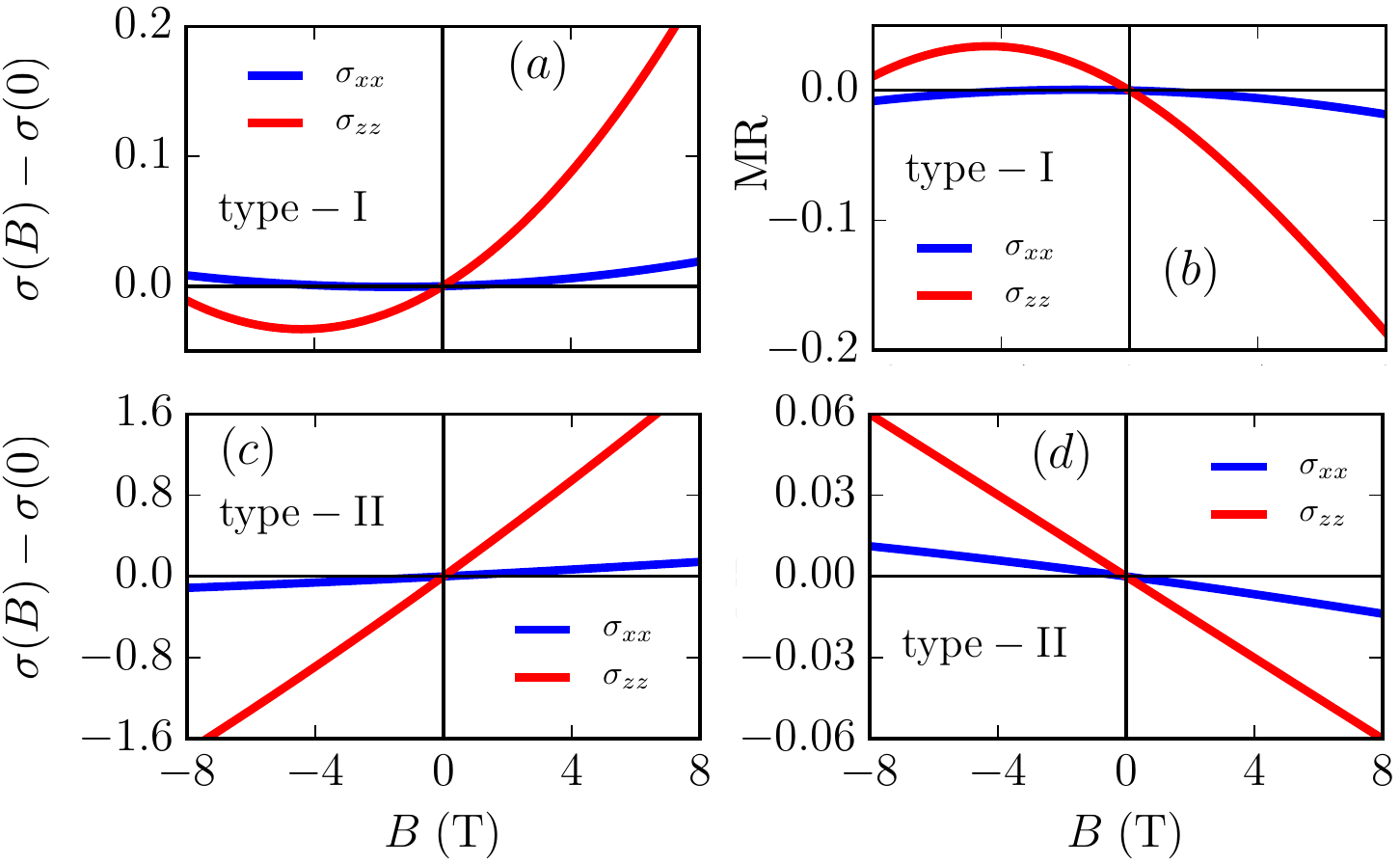}
\caption{The field dependent longitudinal conductivity $[\sigma(B) - \sigma(0) \approx \sigma^{(1)} + \sigma^{(2)}]$ as a function of the field strength for (a) type-I WSM, and 
(c) type-II WSM, respectively for ${\bf B} \parallel \hat{\bf R}$. The corresponding MR [$= \sigma(0)/\sigma(B) - 1$] for a type-I WSM is shown in (b) and for a type-II WSM in (d). The anisotropy in the MR along with the sign change on reversing the direction of $B$,  arising from TRS breaking induced by the tilt in WSM, is evident. Note that while the magnetic component of the conductivity is larger in type-II WSM, on account of the larger density of states at the Fermi energy, the relative MR is smaller in type-II WSM as compared to type-I WSM. All conductivities are in units of $\sigma_{\rm D}$, for ease of comparison. The parameters used for type-I WSM are identical to those in Fig.~\ref{fig:fig_2} and for type-II WSM are identical to those in 
Fig.~\ref{fig:fig_4}.
\label{fig:fig_3}}
\end{figure}

\subsection{Magnetic field perpendicular to the tilt axis (${\bf B} \perp \hat{\bf R}$)}
The $B$-linear terms in the off-diagonal Hall conductivity can be expressed as $\sigma_{yz}^{(1)}=\sigma_{3}\sin\phi$ and $\sigma_{xz}^{(1)}=\sigma_{3}\cos\phi$, where  
\be \label{sigma_3_II_per}
\sigma_{3} = \sum_{s}\dfrac{s\sigma_{1}}{6R_{s}^{4}}
~{\rm sgn}(R_{s})\left[\mathcal{M}
+3\mathcal{F} \left(\delta_{s}^{1}+\delta_{s}^{2}\right)
%
\right].
\ee
Here, we have defined 
\begin{eqnarray}\nn
\mathcal{M}&=&-11+24R_{s}^2-21R_{s}^4~.\\\nn
\end{eqnarray}
Similar to the case of a type-I WSM, these $B$-linear components correspond to the magneto-current of the form  ${\bf j} \propto ({\bf E}\cdot\hat{\bf R}){\bf B}$ (for $\sigma_{xz}^{(1)}$ and $\sigma_{yz}^{(1)}$) or ${\bf j} \propto ({\bf E}\cdot{\bf B}) \hat{\bf R}$ (for $\sigma_{zx}^{(1)}$ and $\sigma_{zy}^{(1)}$).

Now, we discuss the quadratic field dependence of the conductivity matrix. Similar to the case of type-I WSM, the diagonal conductivity components are given by 
$\sigma_{yy}^{(2)} (\phi) = \sigma_{xx}^{(2)} (\pi/2-\phi)$, where we have 
\be \label{sigma_II_xxB2}
\sigma_{xx}^{(2)}(\phi)=\sigma_{4}\cos^2\phi+ \sigma_{5}\sin^2 \phi~.
\ee 
Here, we have defined  the coefficient of cosine term as 
\be
\sigma_{4} = \sum_{s}\dfrac{\sigma_{0}}{16|R_{s}|^5}\left(3-7R_{s}^2+25R_{s}^4+255R_{s}^6+60R_{s}^8\right)~,
\ee
and the coefficient of the sine term as 
\be
\sigma_{5} = \sum_{s}\dfrac{\sigma_{0}}{16|R_{s}|^5}\left(1-5R_{s}^2+15R_{s}^4+5R_{s}^6\right)~.
\ee
Unlike the case of type-I WSM where the coefficient of the sinusoidal term is independent of the tilt parameter $R_s$, 
in Eq.~\eqref{sigma_II_xxB2}, $\sigma_5$ has a significant $R_s$ dependence. The other diagonal component of the quadratic-$B$ conductivity is 
\be \label{sigma_zz_II_2_per}
\sigma_{zz}^{(2)}=\sum_{s}\dfrac{\sigma_{0}}{8|R_{s}|^5}\left(-2+11R_{s}^2-25R_{s}^4+65R_{s}^6+15R_{s}^8\right).
\ee
\begin{table*}
\caption{The BC induced $B$-linear correction to conductivities. Only nonzero corrections to the MC matrix are listed below. The WN separation is assumed to be along the tilt axis, assumed to be $z$ axis in our case. For a WSM with a single pair of oppositely tilted WNs, 
we have, $\sigma^{(1)}_{ij} = 2 \sigma_1~{\rm sign} (R)~\tilde{\sigma}^{(1)}_{ij}$. 
Here $R = R_{+}$ denotes the tilt of the node with positive chirality ($s=+1$), and sign$(R) = +1$ ($-1$) if the $s=+1$ node is tilted in the anticlockwise (clockwise) direction. 
Each entry in the table also references to the corresponding equation number in the main text. \label{T1}
}
\begin{tabular}{cccc}
\hline \hline
\rule{0pt}{5ex} 
\begin{tabular}{c}  $\tilde{\sigma}^{(1)}_{ij} = \tilde{\sigma}^{(1)}_{ji}$ \end{tabular}  
& 
 \begin{tabular}{c} Type-I WSM [$R \to 0+|R|]$ \\ ${\cal{O}}(R)$ \end{tabular}
& 
 \begin{tabular}{c} Type-III WSM [$|R|\to 1-x]$ \\${\cal O}(x)$\end{tabular} 
& 
\begin{tabular}{c}Type-II WSM $[|R|\to 1 + x']$\\$ {\cal O}(x')$\end{tabular}\\[3ex]
\hline
\rule{0pt}{6ex} 
\begin{tabular}{c} $({\bf B}\parallel {\bf R})$\\$\tilde{\sigma}_{xx}^{(1)}=\tilde{\sigma}_{yy}^{(1)}$ \end{tabular}
& 
\begin{tabular}{c} $~\tilde{\sigma}_{zz}^{(1)}\approx  -\frac{46R}{15}$ \eqref{sigma_zz_I_1_par}\\[1ex]
$\tilde{\sigma}_{xx}^{(1)}\approx -\frac{2R}{15}~$\eqref{sigma_xx_I_1_par}
\end{tabular}
&
\begin{tabular}{c}  $\tilde{\sigma}_{zz}^{(1)}\approx\frac{1}{3}(14x-10)$ \eqref{sigma_zz_I_1_par} \\[1ex] 
$\tilde{\sigma}_{xx}^{(1)}\approx -\frac{1}{3}\left[1+\left(7+3\log\frac{x}{2}\right)x\right]$\eqref{sigma_xx_I_1_par}
\end{tabular}  
& 
\begin{tabular}{c} $\tilde{\sigma}_{zz}^{(1)}=-\frac{1}{3}\left[10 + 14x^{\prime}\right]$ \eqref{sigma_zz_II_1_par}\\[1ex]
$\tilde{\sigma}^{(1)}_{xx}\approx -\frac{1}{3}+\left[\frac{13}{3}-\log(2x^{\prime}\Lambda^2)\right]x^{\prime}$ \eqref{sigma_xx_II_1_par}\end{tabular} \\[7ex]
\begin{tabular}{c} 
$({\bf B \perp {\bf R}})$\end{tabular}
& 
\begin{tabular}{c} $\tilde{\sigma}_{yz}^{(1)}=\tilde{\sigma}_{2}\sin\phi,~\tilde{\sigma}_{xz}^{(1)}=\tilde{\sigma}_{2}\cos\phi$ \\[1ex]  where, $\tilde{\sigma}_{2}\approx -\frac{22R}{15}$ \eqref{sigma_2_I_per}\end{tabular}
& 
\begin{tabular}{c} $\tilde{\sigma}_{yz}^{(1)}=\tilde{\sigma}_{2}\sin\phi,~\tilde{\sigma}_{xz}^{(1)}=\tilde{\sigma}_{2}\cos\phi$\\[1ex] where, $\tilde{\sigma}_{2}\approx \frac{1}{3}(2x-4)$ \eqref{sigma_2_I_per}
\end{tabular}  
&
\begin{tabular}{c} $~\tilde{\sigma}_{yz}^{(1)}=\tilde{\sigma}_{3}\sin\phi,~\tilde{\sigma}_{xz}^{(1)}=\tilde{\sigma}_{3}\cos\phi~$
\\[1ex] where, $\tilde{\sigma}_{3}\approx -\frac{1}{3}\left[4 + 2 x^{\prime}\right]$ \eqref{sigma_3_II_per}\end{tabular}\\[4ex]
\hline \hline
\end{tabular}
\end{table*}

The off-diagonal components of the quadratic-$B$ conductivity is given by 
$\sigma_{xy}^{(2)}=\sigma_{6}\sin\phi\cos\phi$, where
\be \label{sigma_xy_II_2_per}
\sigma_{6}=\sum_{s}\dfrac{\sigma_{0}}{8|R_{s}|^5}\left(1-R_{s}^2+5R_{s}^4+125R_{s}^6+30R_{s}^8\right).
\ee
This is the planar Hall conductivity for the case of type-II WSMs. The field angle dependence of the longitudinal conductivity ($\sigma_{xx}^{(2)}$)
and the planar Hall conductivity ($\sigma_{xy}^{(2)}$) for type-II WSM is shown in Figs. \ref{fig:fig_5}(b) and \ref{fig:fig_5}(d) respectively. 
The other off-diagonal quadratic-$B$ terms are zero, similar to the case of type-I WSM, i.e., $\sigma_{xz}^{(2)}=\sigma_{yz}^{(2)}=0$. The tilt dependence of the 
transverse components of the MC for this case, is shown in Fig.~\ref{fig:fig_4}(c). 

\section{Anisotropic Magnetoresistence} 
\label{MR}
In the last two sections, we presented results for the full MC matrix for tilted type-I and ``tilted over" type-II WSM. We show that the TRS breaking induced by the tilt of the WNs in WSM, produces $B$-linear corrections. These $B$-linear terms depend on the sign of the applied magnetic field, and are also sensitive to the direction of the tilt. 
Experimentally these will manifest in the anisotropic nature of the MR measurements \cite{Pengli2017} as a function of the magnetic field, as shown in Fig.~\ref{fig:fig_3}. The magnetic component of the conductivity is larger in type-II WSM, on account of the larger density of states at the Fermi energy as compared to type-I WSM. However, the relative MR turns out to be smaller in type-II WSM as compared to type-I WSM, on account of the relatively larger Drude conductivity. Finally, we emphasize that the MR in both type-I and type-II WSM, can change sign going from positive to negative or vice versa as the sign of $B$ changes. So there is no concrete sign of longitudinal MR either type-I or type-II WSM, in any particular direction, and the actual value depends on various parameters.

\section{The limiting cases: $R_s \to 0$ and $R_s \to 1$}
\label{limiting}
In this section, we summarize all the results obtained in the previous sections, by specifically looking at the three limiting cases of $|R_s| \to 0^+$, $|R_s| \to 1^{-}$ and $|R_s| \to 1^{+}$. All the $B$-linear corrections are summarized in Table~\ref{T1}, while the quadratic-$B$ corrections are summarized in Table~\ref{T2}.  The magneto-current corresponding to the longitudinal $B$-linear terms in the first row of Table~\ref{T1}, can be expressed as ${\bf j} \propto ({\bf R}\cdot{\bf B}) \hat{\bf E}$. On the other hand, 
the transverse $B$-linear terms in the second row of Table~\ref{T1}, correspond either to ${\bf j} \propto ({\bf E}\cdot{\bf B}) \hat{\bf R}$ or to ${\bf j} \propto ({\bf E}\cdot\hat{\bf R}){\bf B}$. 

The correction in the quadratic-$B$ terms are found to be more dominant in the case of ${\bf B} \perp \hat{\bf R}$, as seen from the second row of Table~\ref{T2}. In both type-I and type-II WSM, the quadratic-$B$ components of the conductivity in the {\bf E}-{\bf B} plane for the case of ${\bf B} \perp \hat{\bf R}$, can be expressed as
\bea \label{Eq20x}
\sigma_{xx}&=&\sigma_{\perp}+\Delta\sigma\cos^2\phi~,\\
\sigma_{xy}&=&\Delta\sigma\sin\phi\cos\phi~. \label{Eq21}
\eea
Here, $\Delta\sigma=\sigma_{\parallel}-\sigma_{\perp}$ along with $\sigma_{\parallel}\equiv \sigma_{xx}(\phi=0)$ and $\sigma_{\perp} \equiv \sigma_{xx}(\phi=\pi/2)$.

\begin{table*}
\caption{The BC induced quadratic-$B$ correction to the MC. Only nonzero corrections are listed below. The WN separation is assumed to be along the tilt axis, which is the $z$-axis in our case. For a WSM with a single pair of oppositely tilted WNs, 
we have, $\sigma^{(2)}_{ij} = 2 \sigma_0~\tilde{\sigma}^{(2)}_{ij}$. 
Here $R = R_+$ denotes the tilt of the node with positive chirality ($s=+1$). \label{T2}}
\begin{tabular}{cccc}
\hline \hline
\rule{0pt}{5ex} 
\begin{tabular}{c}  $\tilde{\sigma}^{(2)}_{ij}=\tilde{\sigma}^{(2)}_{ji}$ \end{tabular}  
& 
 \begin{tabular}{c} Type-I WSM [$R \to 0+|R|]$ \\ ${\cal{O}}(R)$ \end{tabular}
& 
 \begin{tabular}{c} Type-III WSM [$|R|\to 1-x]$ \\${\cal O}(x)$\end{tabular} 
& 
\begin{tabular}{c}Type-II WSM $[|R|\to 1 + x']$\\$ {\cal O}(x')$\end{tabular}\\[3ex]
\hline
\rule{0pt}{6ex} 
\begin{tabular}{c}$({\bf B}\parallel {\bf R})$
\end{tabular}
& 
\begin{tabular}{c}  
$~\tilde{\sigma}_{xx}^{(2)}=\tilde{\sigma}_{yy}^{(2)}=1$ [\ref{sigma_xx_I_2_par}]\\[0.5ex]$\tilde{\sigma}_{zz }^{(2)}=8$ \eqref{sigma_zz_I_2_par}
\end{tabular}
&
\begin{tabular}{c} $\tilde{\sigma}_{xx}^{(2)}=\tilde{\sigma}_{yy}^{(2)}=1$
[\ref{sigma_xx_I_2_par}]\\[0.5ex]$\tilde{\sigma}_{zz}^{(2)}=8$ [\ref{sigma_zz_I_2_par}] 
\end{tabular}  
& 
\begin{tabular}{c}$\tilde{\sigma}^{(2)}_{xx}=\tilde{\sigma}^{(2)}_{xx }\approx 1$ [\ref{sigma_xx_II_2_par}]\\[0.5ex]$\tilde{\sigma}_{zz}^{(2)}\approx 8$ \eqref{sigma_zz_II_2_par}\end{tabular} 
\\[7ex]
\begin{tabular}{c}
$({\bf B}\perp {\bf R})$\\$\tilde{\sigma}_{yy}=\tilde{\sigma}_{xx}(\frac{\pi}{2}-\phi)$
\end{tabular} 
&
\begin{tabular}{c}
$\tilde{\sigma}_{zz}^{(2)}\approx 1$ \eqref{sigma_zz_I_2_per}\\ [0.5ex]
$\tilde{\sigma}_{xx}^{(2)}\approx 8\cos^2\phi +\sin^2\phi$ \eqref{sigma_xx_I_2_per}\\[0.5ex]$\tilde{\sigma}_{xy}^{(2)} \approx 7\sin\phi\cos\phi$ \eqref{sigma_xy(2)}
\end{tabular} 
&
\begin{tabular}{c}
$\tilde{\sigma}_{zz}^{(2)}\approx (8-14x)$ \eqref{sigma_zz_I_2_per}\\[1ex]
$~\tilde{\sigma}_{xx}^{(2)}\approx(21-26x)\cos^2\phi + \sin^2\phi$ \eqref{sigma_xx_I_2_per}\\[1ex]
$\tilde{\sigma}_{xy}^{(2)} \approx (10-13x)\sin2\phi$ \eqref{sigma_xy(2)}\end{tabular} 
& 
\begin{tabular}{c}
$\tilde{\sigma}_{zz}^{(2)}\approx (8+ 14x^{\prime})$ \eqref{sigma_zz_II_2_per}\\[1ex]$~\tilde{\sigma}_{xx}^{(2)}\approx(21+26x^{\prime})\cos^2\phi + \sin^2\phi$ \eqref{sigma_II_xxB2}\\[1ex]
$\tilde{\sigma}_{xy}^{(2)} \approx (10+13x^{\prime})\sin2\phi$ \eqref{sigma_xy_II_2_per}
\end{tabular}\\[6ex]
\hline \hline
\end{tabular}
\end{table*}

\section{Impact of internode scattering} \label{internode}
Until now, we have primarily focused on the intranode scattering and ignored the internode scattering. In this section we explore the impact of the internode scattering and chiral anomaly on the MC matrix. This has an impact only for the case of a finite ${\bf E}\cdot{\bf B}$ term. 
We follow the approach of Ref.~[\onlinecite{Yip15}], to calculate the NDF in presence of multiple Fermi surfaces. 
 
To start with, let us assume that initially (say at $t = -\infty$) both the WNs had the same chemical potential of $\mu$. Then, with the onset of 
chiral anomaly induced charge transfer and internode scattering between the nodes, the two WNs eventually acquire a local equilibrium chemical potential (LECP) specified by $\mu_{1/2}$.
Now consider the collision integral for the first node: The intranode scattering will try to keep the LECP at $\mu_1$ while the internode scattering will attempt to force the LECP to be $\mu_2$. 
This can be modelled as, 
\be \label{node_1}
I_{coll}^{(1)}= -\dfrac{g_{\bf k}^{(1)}-f(\epsilon_{\bf k}-\mu_{1})}{\tau_{0}} - \dfrac{g_{\bf k}^{(1)} - f(\epsilon_{\bf k}-\mu_{2})}{\tau_{v}^{12}}~.
\ee 
Here, $\epsilon_{\bf k}$ is the dispersion for node 1, $f(\epsilon_{\bf k} - \mu_i)$ is the Fermi function with Fermi energy $\mu_i$ and $g_{\bf k}^{(i)}$ is the NDF of node $i$. 
Similarly, the other node can have 
\be \label{node_2}
I_{coll}^{(2)}= -\dfrac{g_{\bf k}^{(2)}-f(\epsilon_{\bf k}-\mu_{2})}{\tau_{0}} - \dfrac{g_{\bf k}^{(2)} - f(\epsilon_{\bf k}-\mu_{1})}{\tau_{v}^{21}}~.
\ee 
These equations are supplemented by the equations for particle number conservation. For the case of intranode scattering, we have 
%
\begin{equation}
\sum_{{\bf k} \in 1}\delta {g_{\bf k}}^{(1)}=\sum_{{\bf k} \in 2}\delta {g_{\bf k}}^{(2)}=0~.
\end{equation}
Here, $\delta  g_{\bf k}^{(i)} = g_{\bf k}^{(i)} - f(\epsilon_{\bf k} - \mu_i)$. 
For the case of the internode scattering we have, 
\begin{equation} \label{n_cons2}
\sum_{{\bf k} \in 1}\dfrac{g_{\bf k}^{(1)} - f(\epsilon_{\bf k}-\mu_{2})}{\tau_{v}^{12}} + \sum_{{\bf k} \in 2} \dfrac{g_{\bf k}^{(2)}-f(\epsilon_{\bf k}-\mu_{1})}{\tau_{v}^{21}}=0~.
\end{equation}

To proceed further, let us make the simplifying assumption that the scattering timescale is momentum independent. Then Eq.~\eqref{n_cons2} can be expanded to linear order in $\mu_1 - \mu_2$, 
to obtain
\begin{equation}\label{tau_relation}
\dfrac{\tilde{D}_{1}(B)}{\tau_v^{12}}=\dfrac{\tilde{D}_{2}(B)}{\tau_v^{21}}\equiv\dfrac{\sqrt{D_{1}D_{2}}}{\tau^{X}}~.
\end{equation}
Here, we have defined 
\begin{equation}
\tilde{D}_{i}(B)=\int_{{\bf k}\in i}[d{\bf k}](1+\dfrac{e}{\hbar}{\bf \Omega}_{\bf k}\cdot {\bf B})\left(-\dfrac{\partial f(\mu)}{\partial \epsilon}\right)~,
\end{equation}
and $D_{1/2}$ are simply the density of states of each node in the absence of external fields. The last term of Eq.~(\ref{tau_relation}) defines a field independent internode scattering rate $\tau^X$~.
Using Eqs. (\ref{node_1}) and (\ref{tau_relation}) we can write the collision integral for the first node as
\begin{equation}\label{col_1}
I_{coll}^{(1)}=-\dfrac{\delta g_{\bf k}}{\tilde{\tau}_{1}} - \left(-\dfrac{\partial f(\mu_{1})}{\partial \epsilon}\right) \dfrac{1}{\tau^{X}}\dfrac{\sqrt{D_{1}D_{2}}}{\tilde{D}_{1}(B)}(\mu_{1}-\mu_{2})~.
\end{equation}
Here, we have defined a magnetic field dependent effective scattering timescale, 
\be
\dfrac{1}{\tilde{\tau}_{1}}=\dfrac{1}{\tau_{0}} + \dfrac{1}{\tau^{X}}\dfrac{\sqrt{D_{1}D_{2}}}{\tilde{D}_{1}(B)}~.
\ee
Substituting Eq.~\eqref{col_1} in \eqref{node_1}, we have the following equation for the NDF of the first node, 
\begin{multline}\label{BTE}
\left(1+\dfrac{e}{\hbar}{\bf \Omega}_{\bf k}\cdot {\bf B}\right)^{-1}\left(e{\bf E} + \dfrac{e^2}{\hbar} ({\bf E}\cdot{\bf B})\right)\cdot\dfrac{\partial \epsilon_{\bf k}}{\partial {\bf k}}\left(-\dfrac{\partial f(\mu_1)}{\partial \epsilon}\right)
\\
=-\dfrac{\delta g_{\bf k}}{\tilde{\tau}_{1}}
-\dfrac{\delta \mu_{1}-\delta \mu_{2}}{\tau^{X}}\dfrac{\sqrt{D_{1}D_{2}}}{\tilde{D}_{1}(B)}\left(-\dfrac{\partial f(\mu_1)}{\partial \epsilon}\right)~,
\end{multline}
where we have defined $ \delta \mu_{1/2} = \mu_{1/2} - \mu$, with $\mu$ being the initial chemical potential without any charge imbalance. 
Now the charge current can be written as
\be \label{current}
{\bf j}^e = -e\int [d{\bf k}]\tilde{v}_{\bf k} 
\left[\delta g_{\bf k} + \delta \mu_{1}
\left(-\partial_\epsilon f(\mu_1)\right)\right]~.
\ee
with $\tilde{v}_{\bf k} \equiv {\bf v}_{\bf k} + \dfrac{e}{\hbar}({\bf v}_{\bf k} \cdot {\bf \Omega}_{\bf k}){\bf B}$.

Explicit forms of $\delta g_{\bf k}$ and $\delta \mu_{1/2}$ are yet to be obtained. To this end, 
charge conservation forces the following relation 
\be 
\tilde{D}_{1}(B) \delta \mu_{1} + \tilde{D}_{2}(B) \delta \mu_{2} = 0~. 
\ee
For WSM we can show that $\tilde{D}_{1}(B)=\tilde{D}_{2}(B)$ and hence $\delta \mu_{1}= -\delta \mu_{2}$. 
From Eq. \eqref{BTE} taking $\sum_{\bf k}$ on both sides we get the identity
\be \label{chrl_anmly}
\delta{\mu}_{1}^2 - \delta{\mu}_{2} 
=-\dfrac{e\tau^{X}}{\sqrt{D_{1}D_{2}}} \int [d{\bf k}]
(\tilde{\bf v}_{\bf k} \cdot {\bf E})
\left(-\dfrac{\partial f(\mu_{1})}{\partial \epsilon}\right)~.
\ee
This is a generalized form of the chiral anomaly induced chemical potential difference of the Weyl nodes, including the impact of both internode scattering as well as 
intranode scattering. 
Now using Eq.~\eqref{chrl_anmly} in \eqref{BTE}, $\delta g_{\bf k}$, is found to be 
\begin{multline} \label{NDF}
\delta g_{\bf k}=\bigg[-\tilde{\tau}_{1}\left(1+\dfrac{e}{\hbar}{\bf \Omega}_{\bf k} \cdot {\bf B}\right)^{-1}
\left(e{\bf E} + \dfrac{e^2}{\hbar} ({\bf E}\cdot{\bf B}){\bf \Omega}_{\bf k}\right) \cdot {\bf v}_{\bf k}
\\
-2\delta \mu_{1}\dfrac{\tilde{\tau}_{1}}{\tau^{X}}\dfrac{\sqrt{D_{1}D_{2}}}{\tilde{D}_{1}(B)}\bigg]
\left(-\dfrac{\partial f(\mu_{1})}{\partial \epsilon}\right)~.
\end{multline}
For linear response we can expand the derivative of Fermi function (in the last term of Eq.~\eqref{NDF}) around the equilibrium chemical potential ($\mu$) because the $\delta \mu_{1/2}$ itself is a function of electric field.
Finally,  the current in Eq. \eqref{current} can be expressed as  
\begin{multline}\label{current_f}
{\bf j}^e = e^{2}\tilde{\tau}_{1}\int [d{\bf k}]
\left(1+\dfrac{e}{\hbar}{\bf \Omega}_{\bf k}\cdot {\bf B}\right)^{-1} \tilde{\bf v}_{\bf k}(\tilde{\bf v}_{\bf k}\cdot {\bf E})
\left[- f^{\prime}\right]
\\
-e\int [d{\bf k}] \tilde{\bf v}_{\bf k}
\left[- f^{\prime}\right]
\delta \mu_{1}
\left(1 - 2 \dfrac{\tilde{\tau}_{1}}{\tau^{X}}\dfrac{\sqrt{D_{1}D_{2}}}{\tilde{D}_{1}(B)}\right)~.
\end{multline}
\
Here the first term is dominated by the intranode scattering which has been discussed in detail in this paper. 
The second term $\propto \delta \mu_1$, is chiral anomaly and internode scattering dominated. This becomes more clear by substituting the value of $\delta{\mu}_{1}$ in the second term of Eq.~\eqref{current_f}, and expressing it as a sum of three terms ${\bf j}^e = {\bf j}_1^e + {\bf j}_2^e + {\bf j}_3^e$. 
The contribution which is dominated by the intranode scattering reads as
\begin{equation}\label{current_1}
{\bf j}_{1}^e = e^{2}\tilde{\tau}_{1}\int [d{\bf k}]
\left(1+\dfrac{e}{\hbar}{\bf \Omega}_{\bf k}\cdot {\bf B}\right)^{-1}\tilde{\bf v}_{\bf k}(\tilde{\bf v}_{\bf k}\cdot {\bf E})
\left[- f^{\prime}\right].
\end{equation}
We have used this expression of current for the detailed calculation in this paper.
The contribution to current arising from chiral anomaly and dominated by the internode scattering, is 
\be \label{current_2}
{\bf j}_2^e =\dfrac{e^{2}\tau^{X}}{2\sqrt{D_{1}D_{2}}}
\left[\int [d{\bf k}] \tilde{\bf v}_{\bf k}
\left[-f^{\prime}\right]\right]
\left[\int[d{\bf k}](\tilde{\bf v}_{\bf k}\cdot {\bf E})
\left[- f^{\prime}\right]\right]~.
\ee
This term is similar in spirit to the chiral anomaly induced negative MR, 
discussed first in Ref.~[\onlinecite{Son_Spivak2013}]. 
The third term is also chiral anomaly induced, but is 
dominated by the intranode scattering: 
\be \label{current_3}
{\bf j}_{3}^e=-\dfrac{e^{2}\tilde{\tau}_{1}}{\tilde{D}_{1}(B)}
\left[\int [d{\bf k}] \tilde{\bf v}_{\bf k}
\left[-f^{\prime}\right]\right]
\left[\int[d{\bf k}](\tilde{\bf v}_{\bf k}\cdot {\bf E})
\left[- f^{\prime}\right]\right]~.
\ee
This summarizes the semiclassical Boltzmann transport formalism in systems with multiple Fermi surfaces. 

To explore the impact of the chiral-anomaly and the associated internode timescale on the MC we consider an isotropic WSM without tilt. 
The density of states is given by $D_{1}=D_{2}={\mu^{2}}/({2 \pi^2 \hbar^3 v_{F}^3)}$.
The LECP of the WNs is calculated from Eq.~\eqref{chrl_anmly} and it is given by 
\be 
\mu_{s}= \mu + s\dfrac{e^2 \hbar v_{F}^3}{2\mu^2}({\bf E}\cdot {\bf B}) \tau_{v},
\ee
where we have used $\tau_{v}=\tau^{X}/2$, and $s$ denotes the chirality of each node. 
This $({\bf E}\cdot {\bf B}) \tau_{v}$ dependent difference in chemical potential induced by the chiral anomaly, also manifests in the longitudinal and planar transport properties. 

For the case when both ${\bf E}$ and ${\bf B}$ are in along the $z$-axis, the internode scattering induced longitudinal conductivity can be obtained from Eq.~\eqref{current_2}, and it is given by 
\be \label{sigma_CA}
\sigma_{zz}^{(2)}=\dfrac{e^{2}}{4\pi^2 \hbar}\dfrac{(eB)^2 v_{F}^{3}}{\mu^2}\tau_{v}~.
\ee

This term contributes to the chiral anomaly induced negative MR, and is identical to the result of Ref.~[\onlinecite{Son_Spivak2013}]. 
For the other case of a co-planar setup (a planar-Hall geometry), 
a finite contribution to the planar Hall effect can be obtained. From Eq.~\eqref{current_2} we calculate 
\be
\sigma_{xx}^{(2)} = \Delta \sigma^{\rm CA} \cos^2 \phi~,~~~
\sigma_{xy}^{(2)} = \Delta \sigma^{\rm CA} \sin \phi \cos \phi~.
\ee
Here $\Delta \sigma^{\rm CA}$ is equal to $\sigma_{zz}$ defined in Eq. \eqref{sigma_CA}.
This is the chiral anomaly induced planar Hall conductivity in WSM, which has also been derived earlier \cite{Burkov2017}. 
Finally, we note that the impact of the chiral anomaly will be observable only if $3e^2\hbar v_{F}^3{\bf E\cdot B}~\tau_v/{(2 \mu^3)} \approx 1$. For reasonable values of $\tau_v = 10^{-9}$ s and $v_F = 10^6$ m/s, this results in ${\bf E \cdot B} \approx \mu^3 \times 10^6$ Tesla V/m, when $\mu$ is in eV.

\section{Conclusion}
\label{Concl} We have presented a systematic analysis of the impact of the TRS breaking tilting of the WNs for both type-I and type-II WSM, on the MC. 
While several earlier works have calculated the conductivity matrix elements for specific cases, we calculate the 
full conductivity matrix, explicitly for the two different cases of the applied magnetic field being parallel to the tilt axis, and perpendicular to it.
In addition to the Drude conductivity, and the quadratic-$B$ terms of the conductivity matrix which have been studied earlier, we also show the existence of 
previously unexplored $B$-linear terms in the conductivity matrix. Our calculations also include the impact of the tilt of the type-I as well as type-II Weyl nodes to all orders, analytically. 
One limitation of our approach is that we have not modelled the anisotropic nature of the intranode and internode node scattering timescales, and consider both to be a constant for a given Fermi surface.

We find that the BC and the TRS breaking tilt of the WSM, combine to produce $B$-linear anisotropic corrections in MC (see Table~\ref{T1} for a quick summary). 
In particular we predict the previously unexplored, out-of-plane transverse conductivity ($\sigma_{xz}^{(1)} = \sigma_{zx}^{(1)}$ and $\sigma_{yz}^{(1)}=\sigma_{zy}^{(1)}$) in tilted WSM when $B$ is  applied perpendicular to the tilt axis. For these terms the corresponding magneto-current is either of the form ${\bf j} \propto ({\bf E}\cdot{\bf B}) \hat{\bf R}$ or ${\bf j} \propto ({\bf E}\cdot\hat{\bf R}){\bf B}$, with the proportionality constant being a non-linear function of the tilt (parametrized by $R$). 
Interestingly, these $B$-linear conductivity terms depend on the chemical potential only through $\tau_\mu$, while being sensitive to the tilt orientation and the magnetic field direction.  For the case of $B$ applied along the tilt axis, we find that the longitudinal conductivities also get $B$-linear corrections. For these corrections terms we have,  
${\bf j} \propto ({\bf B}\cdot\hat{\bf R}) {\bf E}$. In experiments, this $B$-linear MC correction will result in anisotropic MR which also flips sign with reversal of $B$, as indicated in Fig.~\ref{fig:fig_4}. 

In addition to the $B$-linear corrections, the TRS breaking tilt of the WNs also modifies the quadratic terms as well. These quadratic-$B$ corrections to the MC are summarized in Table~\ref{T2}. 
For the case when $B$ is applied along the tilt direction, the diagonal terms of the MC matrix get tilt dependent quadratic-$B$ corrections (see row 1 of Table~\ref{T2}). For the case when $B$ is applied perpendicular to the tilt axis, we generalize the earlier known results for the longitudinal quadratic-$B$ conductivity to include the correction arising from the phase factor correction, and the impact of the tilt. 
We also affirm the existence of planar Hall effect in tilted WSM with the same angular dependence as in the isotropic case \cite{Nandy2017} $\sigma_{xy} \propto \sin(2\phi)$, although the proportionality constant depends on the degree of the tilt as highlighted in the second row of Table~\ref{T2}. In addition to this, we also find a finite quadratic-$B$ correction in the MC arising from the phase space terms, which has been missed in earlier papers. This will manifest in the planar Hall conductivity measurements in WSM \cite{Nitesh2018}, and it modifies the known relation, $\rho_{xx}=\rho_{\perp}-\Delta\rho\cos^2\phi$, making $\rho_{\perp}$ a field dependent quantity.

\appendix{

\section{The limiting case of $Q \to 0$ in type-II WSM}
\label{appendixA}
For the case of type-II WSM, the ultraviolet cutoff scheme can also be introduced by using a spherical geometry along with a radial momentum cut-off. 
This is particularly useful for the case of both WNs being situated at the origin, i.e, the $Q \to 0$ limit of Eq.~\eqref{Ham_type_I}:  
\be\label{ham}
\mathcal{H}_{s}({\bf k}) = \hbar C_{s}k_{z}+s\hbar v_{F} {\bf \sigma}\cdot{\bf k}~.
\ee
This cutoff scheme preserves spherical symmetry, and allows the Brillouin zone integrations to be done using polar coordinates. Accordingly, the momentum cutoff ($\Lambda_k$) can be chosen to be in the radial direction. This allows us to check the consistency of all our calculations.

In zero magnetic field, we find the diagonal components of the conductivity matrix to be $\sigma_{xx}^{(0)}=\sigma_{yy}^{(0)}$, where
\be\label{sigma_xx_0_ori_II}
\dfrac{\sigma_{xx}^{(0)}}{\sigma_{\rm D}}=\sum_{s}\dfrac{3}{4|R_{s}|^{3}}\left[\dfrac{3-R_{s}^{2}}{R_{s}^{2}-1}-\delta_{s}^{1}
+ \mathcal{F}^{1/2}\tilde{\Lambda}_k^{2}
-2~\ln\tilde{\Lambda}_k\right].
\ee
Here $\tilde{\Lambda}_k = \Lambda_k/k_F$. As an additional consistency check we note that Eq.~\eqref{sigma_xx_0_ori_II} is consistent with Eq.~(30) in this paper, and Eq.~(B3) in Ref.~[\onlinecite{carbotte}].  

 The Drude conductivity along the tilt ($\hat{z}$-) axis is given by
\be\label{sigma_zz_0_ori_II}
\dfrac{\sigma_{zz}^{(0)}}{\sigma_{\rm D}}=\sum_{s}\dfrac{3}{2|R_{s}|^{3}}\left[3-R_{s}^{2}+\delta_{s}^{1}
+ \mathcal{F}^{1/2}\tilde{\Lambda}_k^{2}+2~\ln\tilde{\Lambda}_k\right]~.
\ee

\subsection{Magnetic field along the tilt axis (${\bf B} \parallel \hat{\bf R} $)}
The $B$-linear correction to the diagonal component of the conductivity matrix is given by 
\be\label{sigma_zz_1_II_par}
\sigma_{zz}^{(1)}=\sum_{s}\dfrac{s\sigma_{1}}{3R_{s}^{4}}{\rm sgn}(R_{s})\left(\mathcal{A}_{z}-3\mathcal{F}\delta_{s}^{1}
-6\mathcal{F}~\ln \tilde{\Lambda}_k\right)~.
\ee
Unlike the case of type-I WSM, in Eq.~\eqref{sigma_zz_1_II_par}, there is Fermi energy dependence that is tied to the cutoff. The total linear-$B$ contribution of two nodes having parallel tilt, becomes zero as in the type-I case. But for nodes having opposite tilt, as is usually the case, the total contribution is non-zero. The sign of the $B$-linear terms is sensitive to the relative orientation of the tilt. The $B$-linear correction to the other diagonal components are given by $\sigma_{xx}^{(1)}=\sigma_{yy}^{(1)}$, where
\be\label{sigma_xx_1_II_par}
\sigma_{xx}^{(1)}=\sum_{s}\dfrac{s\sigma_{1}}{6R_{s}^{4}}{\rm sgn}(R_{s})\left(\mathcal{A}_{x}-3\mathcal{F}^{1/2}\delta_{s}^{1}
-6\mathcal{F}^{1/2}~\ln \tilde{\Lambda}_k\right)~.
\ee
The $B$-linear corrections to the all the transverse components for this configuration are identically zero, similar to the case of type-I WSM (see Table~\ref{T1}).

The quadratic-$B$ dependence of the MC along the tilt direction is given by 
\be\label{sigma_zz_2_II_par}
\sigma_{zz}^{(2)}=\sum_{s}\dfrac{\sigma_{0}}{2|R_{s}|^{5}}\left(1-5R_{s}^{2}+15R_{s}^{4}+5R_{s}^{6}\right)~.
\ee
Surprisingly, the quadratic-$B$ correction has no cutoff dependence unlike the Drude and the $B$-linear conductivities. 
The quadratic correction to the other diagonal elements is given by $\sigma_{xx}^{(2)}=\sigma_{yy}^{(2)}$, where 
\be\label{sigma_xx_2_II_par}
\sigma_{xx}^{(2)}=\sum_{s}\dfrac{\sigma_{0}}{8|R_{s}|^{5}}
\left(-2+5R_{s}^{2}+5R_{s}^{6}\right)~.
\ee
The quadratic-$B$ correction to all the transverse conductivity components is identically zero, similar to the case of type-I WSM.

\subsection{Magnetic field perpendicular to the tilt axis (${\bf B} \perp \hat{\bf R}$)}
Unlike the case of type-I WSM,  the out-of-plane transverse conductivities have a $B$-linear component given by  $\sigma_{xz}^{(1)}=\sigma_{2}\cos\phi$ and $\sigma_{yz}^{(1)}=\sigma_{2}\sin\phi$, where we have defined
\be
\sigma_{2}=\sum_{s}\dfrac{s\sigma_{1}}{6R_{s}^{4}}{\rm sgn}(R_{s})\bigg(\mathcal{M} 
+3\mathcal{F}\delta_{s}^{1}+6\mathcal{F}~\ln\tilde{\Lambda}_k\bigg)~.
\ee
The $B$-linear dependence of the rest of the conductivities is zero, i.e. $\sigma_{xy}^{(1)}=\sigma_{xx}^{(1)}=\sigma_{yy}^{(1)}=\sigma_{zz}^{(1)}=0$.

The quadratic-$B$ dependence of in-plane component of longitudinal conductivities is given by $\sigma_{xx}^{(2)}=\sigma_{3}\cos^{2}\phi+\sigma_{4}\sin^{2}\phi$, where we have defined
\be
\sigma_{3} = \sum_{s}\dfrac{\sigma_{0}}{16|R_{s}|^{5}}
\left(3-7R_{s}^{2}+25R_{s}^{4}+255R_{s}^{6}+60R_{s}^{8}\right)~,
\ee
and
\be
\sigma_{4} = \sum_{s}\dfrac{\sigma_{0}}{16|R_{s}|^{5}}\left(1-5R_{s}^{2}+15R_{s}^{4}+5R_{s}^{6}\right)~.
\ee
Recall that $\sigma_{yy}^{(2)}(\phi) = \sigma_{xx}^{(2)}(\pi/2-\phi)$. So the coefficients of trigonometric function gets interchanged. The quadratic-$B$ component of the other diagonal term of the conductivity matrix, is given by 
\be \label{sigma_zz_2_II_per}
\sigma_{zz}^{(2)}= \sum_{s}\dfrac{\sigma_{0}}{8|R_{s}|^{5}}\left(-2+11R_{s}^{2}-25R_{s}^{4}+65R_{s}^{6}+15R_{s}^{8}\right)~.
\ee
The in-plane Hall conductivity has quadratic-$B$ dependence and it is given by $\sigma_{xy}^{(2)}=\sigma_{5}\sin\phi\cos\phi$, where
\be \label{sigma_xy_2_II_per}
\sigma_{5}=\sum_{s}\dfrac{\sigma_{0}}{8|R_{s}|^{5}}\left(1-R_{s}^{2}+5R_{s}^{4}+125R_{s}^{6}+30R_{s}^{8}\right)~.
\ee
The quadratic dependence of other transverse conductivities are zero, i.e. $\sigma_{xz}^{(2)}=\sigma_{yz}^{(2)}=0$.

\section{Details of the calculations using cylindrical geometry}  
\label{appendixB}
For a pair of WNs separated in the momentum space, the conductivities can be calculated by exploiting the cylindrical symmetry of the system.
The general expression for different components of the MC (at $T = 0$) can be expressed as 
\begin{widetext}
\be \label{sigma_xx_0_sep}
\sigma=\dfrac{e^{2}\tau_\mu}{(2\pi)^{3}}
\int_{0}^{\infty}k_{\perp}dk_{\perp}\int_{0}^{2\pi}d\phi_{\bf k}\int_{-\Lambda_{z}}^{{\Lambda_{z}}}dk_{z}~
\mathcal{L}(k_{\perp}, \phi_{\bf k}, k_{z}-s Q )~
\delta\left(\mu-\hbar \left[C_{s}(k_{z}-s Q)\pm v_{F}\sqrt{k_{\perp}^{2}+(k_{z}-s Q)^{2}}\right]\right).
\ee
\end{widetext}
Here $k_{\perp}^2=k_{x}^2+k_{y}^2$ and $\mathcal{L}(k_{\perp}, \phi_{\bf k}, k_{z}-s Q )$ is an unspecified function.
The integral in Eq.~\eqref{sigma_xx_0_sep}, can be simplified using the following property of the Dirac delta function:
\be  \label{root_eqn}
\delta[f(x)]=\sum_i\dfrac{\delta(x-x_{i})}{|f^{\prime}(x_{i})|}~,
\ee
where $x_{i}$ are the zeros of $f(x)$. For the conduction band ($\lambda = +$), the zeros of $f(k_{\perp})$ are given by 
\be\label{zero_c} 
k_{\perp}^{2}=\left(k_{F}-R_{s}\tilde{k}_{z,s}\right)^{2}-\tilde{k}_{z,s}^{2}~,
\ee
where $\tilde{k}_{z,s}\equiv (k_{z}-sQ)$. The same for the valance band are 
\be\label{zero_v} 
k_{\perp}^{2}=\left(R_{s}\tilde{k}_{z,s}-k_{F}\right)^{2}-\tilde{k}_{z,s}^{2}~.
\ee
In both these expressions we have assumed $\mu>0$ and have defined $k_{F}=\frac{\mu}{\hbar v_{F}}$. 
From Eq. \eqref{zero_c} we get 
\be
\tilde{k}_{s}=\left(k_{F}-R_{s}\tilde{k}_{z,s}\right)~,
\ee
where $\tilde{k}_{s}^2\equiv k_{\perp}^2+\tilde{k}_{z,s}^2$. As $\tilde{k}_{s} \ge 0$ we must have have 
$\tilde{k}_{z,s} \leq \frac{k_{F}}{R_{s}}$ for positive $R_{s}$, and $\tilde{k}_{z,s} \geq \dfrac{k_{F}}{R_{s}}$ for negative $R_{s}$. Similarly for the 
the valence band, from Eq. \eqref{zero_v}, we get exactly opposite conditions, $\tilde{k}_{z,s} \geq \frac{k_{F}}{R_{s}}$ for positive $R_{s}$ and $\tilde{k}_{z,s}\leq  \frac{k_{F}}{R_{s}}$ for negative $R_{s}$. We emphasize that these conditions are independent of the absolute value of $R_s$ and are valid for both type-I and type-II WSM. In the following subsections we explicitly calculate the different limits of integration for type-I and for type-II WSM.

\subsection{Type-I}
For a type-I Weyl node ($|R_{s}|<1$), Eq.~\eqref{zero_c} can be expressed as 
\be\label{zero_c_I} 
k_{\perp}=\left[\left(k_{F}+(1- R_{s})\tilde{k}_{z,s}\right)\left(k_{F}-(1+R_{s})\tilde{k}_{z,s}\right)\right]^{1/2}.
\ee
Since $k_{\perp}$ is a real number, for $R_s > 0$, this restricts $\tilde{k}_{z,s}$ to be 
\be 
-\dfrac{k_{F}}{1-R_{s}} \leq \tilde{k}_{z,s} \leq \dfrac{k_{F}}{1+R_{s}}~.
\ee
Note that this condition does not violate the condition imposed on $\tilde{k}_{z,s}$ for $\tilde{k}_{s}$ to be positive. For $R_{s}<0$, 
Eq.~\eqref{zero_c_I} leads to 
\be 
-\dfrac{k_{F}}{1+|R_{s}|} \leq \tilde{k}_{z,s} \leq \dfrac{k_{F}}{1-|R_{s}|}~.
\ee
Using these limits of integration for the cylindrical geometry, we get results which are identical to the ones obtained 
by using spherical geometry with both the nodes being centered at the origin. 

\subsection{Type-II}
For type-II WN, the zero of Dirac delta function for the conduction band in Eq. \eqref{zero_c},  can be expressed as 
\be\label{root_type_II} 
k_{\perp}=\left[\left(k_{F}-(R_{s}-1)\tilde{k}_{z,s}\right)\left(k_{F}-(R_{s}+1)\tilde{k}_{z,s}\right)\right]^{1/2}~.
\ee
The value of the tilt factor makes this expression different from Eq.~\eqref{zero_c_I}.  
Note that unlike the type-I WSM, the real nature of $k_{\perp}$ does not force $\tilde{k}_{z,s}$ to be bounded as both the terms in Eq.~\eqref{root_type_II} are positive (negative) for large negative (positive) value of $\tilde{k}_{z,s}$. Hence we are forced to choose a momentum cutoff ($\Lambda_z$). 
With the cutoff, the appropriate integration limit for the case of $R_{s}>1$ is given by 
\be \label{limit_c_pos} 
-(\Lambda_{z}+sQ) \leq \tilde{k}_{z,s} \leq \dfrac{k_{F}}{R_{s}+1}~.
\ee
For the case of $R_{s}<-1$ we get 
\be \label{limit_c_neg} 
-\dfrac{k_{F}}{|R_{s}|+1} \leq \tilde{k}_{z,s} \leq \Lambda_{z}-sQ~.
\ee
Note that these limits are significantly different when compared to that of type-I WSM. 
For the valence band, Eq. \eqref{zero_v} can be re-written as
\be\label{root_type_II} 
k_{\perp}=\left[\left((R_{s}-1)\tilde{k}_{z,s}-k_{F}\right)\left((R_{s}+1)\tilde{k}_{z,s}-k_{F}\right)\right]^{1/2}~.
\ee
Here both the factors inside the root are positive (negative) for large positive (negative) value of $\tilde{k}_{z,s}$. Hence, choosing a finite cutoff as done for the conduction band, the limits of the integration for $R_{s}>1$ is calculated to be
\be \label{limit_v_pos} 
\dfrac{k_{F}}{R_{s}-1} \leq \tilde{k}_{z,s} \leq \Lambda_{z}-sQ~.
\ee
Note that for $R_s >0$, for the conduction band,  $\tilde{k}_{z,s}$ varies from a positive value to negative, whereas for the valence band integration, the limits lie only on the positive axis of $\tilde{k}_{z,s}$.
For $R_{s}<-1$, instead of Eq. \eqref{limit_v_pos}, we get the finite contribution in the limit 
\be \label{limit_v_neg} 
-(\Lambda_{z}+sQ) \leq \tilde{k}_{z,s} \leq -\dfrac{k_{F}}{|R_{s}|-1}~.
\ee
Equations \eqref{limit_c_pos}-\eqref{limit_v_neg} are the main results of this section. These limits are essential to calculate all the quantities of interest for type-II WSM.

\section{Details of the calculation using spherical geometry}  
\label{appendixC}
An alternate way of doing all the calculations, is to use the $Q \to 0$ limit along with a spherical geometry with a radial cutoff as done in Ref.~[\onlinecite{carbotte}]. 
The general expression for conductivities in this case, is given by
%
\bea  \label{sigma_xx_0_sep_c2}
\sigma &=&\dfrac{e^{2}\tau_\mu}{(2\pi)^{3}}
\int_{0}^{\infty}k^2dk\int_{0}^{2\pi}d\phi_{\bf k}\int_{0}^{\pi}\sin\theta_{\bf k}d\theta_{\bf k}~
\mathcal{L}(k, \theta_{\bf k},\phi_{\bf k})~\nn \\
& & \times \delta\left(\mu-\hbar \left[C_{s}k\cos\theta_{\bf k}\pm v_{F}k\right]\right)~.
\eea
After changing variable, $\cos \theta_{\bf k} \to x$, it can be rewritten as 
\bea \label{sigma_xx_0_sep_c}
\sigma &=&\dfrac{e^{2}\tau_\mu}{(2\pi)^{3}}
\int_{0}^{\infty}k^2dk\int_{0}^{2\pi}d\phi_{\bf k}\int_{-1}^{1}dx~
\mathcal{L}(k, x,\phi_{\bf k})~\nn \\
& & \times \delta\left(\epsilon_{F}-\hbar \left[C_{s}kx\pm v_{F}k\right]\right)~.
\eea
%
As usual, to solve this integration we use roots of the Dirac delta function as in Eq.~\eqref{root_eqn}. 
In this case the roots for the conduction band can be calculated to be  
\be
k_{r}=\dfrac{k_{F}}{R_{s}x+1}~.
\ee
For the positivity of $ k_r$ we must have $(R_{s}x+1)>0$. 
Similarly, for the valence band it is given by
\be 
k_{r}=\dfrac{k_{F}}{R_{s}x-1},
\ee
along with the condition that $(R_{s}x-1)>0$. It can be easily checked that for type-I WSM, $x$ varies over its entire range $-1$ to $1$, while it is not so for the case of 
a type-II WSM. 

For a type-II WSM,  $x=-1/R_{s}$ sets the limit of $k$ to infinity. To remedy this situation we introduce a finite momentum cut-off $\Lambda_k$ such that $\Lambda_k (1 + R)>k_{F}$. For the conduction band with $R_{s}>1$ we obtain the integration limit to be 
\be
\left(\dfrac{k_{F}}{\Lambda_k}-1\right)\dfrac{1}{R_{s}} \leq x  \leq 1~,
\ee
with $\Lambda_k(R_{s}+1)>k_{F}$. The term on the left hand side of this inequality is negative for $R_s >0$. When $R_{s}<-1$, we get
\be
-1 \leq x  \leq \left(\dfrac{k_{F}}{\Lambda_k}-1\right)\dfrac{1}{R_{s}}~,
\ee
with $\Lambda_k(-R_{s}+1)>k_{F}$. Similarly, for the valence band we get the following limit 
\be
\left(\dfrac{k_{F}}{\Lambda_k}+1\right)\dfrac{1}{R_{s}} \leq x \leq 1~,
\ee
with $\Lambda_k(R_{s}-1)>k_{F}$ for $R_{s}>1$. For $R_{s}<-1$ we get the following limit
\be 
-1 \leq x \leq \left(\dfrac{k_{F}}{\Lambda_k}+1\right)\dfrac{1}{R_{s}},
\ee
with $-\Lambda_k(R_{s}+1)>k_{F}$. These limits are essential for performing all the integrals for calculating the conductivities.

\section{Exact results for type-II WSM}\label{Appendix-type-II}
In the main text we have presented results up to first order in $Q/\Lambda_z$ for the sake of simplicity and ease of understanding, even though all our calculations 
retain all orders in $Q/\Lambda_z$. Thus for completeness, we present the exact results here.  
The diagonal components of the exact Drude conductivity matrix, is given by $\sigma_{xx}^{(0)}=\sigma_{yy}^{(0)}$, where
\begin{multline} \label{sigma_xx_0_sep_II_c2}
\sigma_{xx}^{(0)} = \frac{3\sigma_{\rm D}}{4}\sum_{s}\bigg[\dfrac{1}{|R_{s}|}
\bigg(\dfrac{R_{s}^2+1}{R_{s}^2-1}-\dfrac{\delta_{s}^{1}}{R_{s}^{2}}\bigg)+
2s\tilde{Q}\dfrac{R_{s}^2+1}{R_{s}^{2}}{\rm sgn}(R_{s})
\\
+\dfrac{R_{s}^{2}-1}{|R_{s}|}~\Gamma
-\dfrac{\delta_{s}^{3}}{|R_{s}|^{3}}\bigg]~.
\end{multline}
Here we have defined $\Gamma \equiv (\Lambda_{z}^2+Q^2)/k_{F}^2$ and $\tilde{Q} \equiv Q/k_{F}$. The quantity $\delta_{s}^{1}$ is defined earlier and $\delta^{3}_{s} \equiv \ln\left[R_{s}^2\gamma-1-2s\tilde{Q}R_{s}\right]$, along with $\gamma = (\Lambda_{z}^2-Q^2)/k_{F}^2$. 
The other diagonal or Drude component of conductivity along the tilt direction is 
\begin{multline}
\sigma_{zz}^{(0)} =\dfrac{3\sigma_{\rm D}}{2}\sum_{s} \bigg[\dfrac{1}{|R_{s}|}\bigg(
R_{s}^2+1+\dfrac{\delta_{s}^{1}}{R_{s}^{2}}\bigg)
+2s\tilde{Q}\dfrac{R_{s}^4-1}{R_{s}^2}~{\rm sgn}(R_{s})
\\
+\dfrac{(R_{s}^{2}-1)^2}{|R_{s}|}~\Gamma
+\dfrac{\delta_{s}^{3}}{|R_{s}|^{3}}\bigg]~.
\end{multline}
The off-diagonal components of zero field conductivities are identically zero, i.e., $\sigma_{xy}^{(0)}=\sigma_{xz}^{(0)}=\sigma_{yz}^{(0)}=0$.

\subsection{Magnetic field along the tilt (${\bf B} \parallel \hat{\bf R}_{s}$)}
In this section we discuss the magneto-transport when magnetic field is parallel to the tilt.
In this scenario we get linear as well as quadratic magnetic field correction in the diagonal components of conductivities. The $B$-linear correction to diagonal component along the tilt axis is given by
\begin{multline}
\sigma_{zz}^{(1)}=\sum_{s}\dfrac{s\sigma_{1}}{6R_{s}^{4}}~{\rm sgn}(R_{s})\Big[2\mathcal{A}_{z}-6\mathcal{F}\left(\delta_{s}^{1} + \delta_{s}^{3}\right)
\\
-2\left(p^{3} - q^{3}\right)+9\left(p^2 + q^{2}\right)+6\mathcal{G}\left(p-q\right)\Big]~.
\end{multline}
Here, we have defined the following
\bea\nn
p&=&\left(|R_{s}|[\tilde{\Lambda}_{z}+ s\tilde{Q}~ {\rm sgn}(R_s)]+1\right)^{-1}~,\\\nn
q&=&\left(|R_{s}|[\tilde{\Lambda}_z - s\tilde{Q}~{\rm sgn}(R_s)]-1\right)^{-1}~.
\eea
Other polynomials of $R_{s}$, i.e., $\mathcal{A}_{1}^{z}$ and $\mathcal{F}$ are defined earlier and we define
$\mathcal{G}\equiv(R_{s}^4+2R_{s}^2-3)$.
The linear-$B$ correction to the other diagonal components is given by $\sigma_{xx}^{(1)}=\sigma_{yy}^{(1)}$, where
\begin{multline}
\sigma_{xx}^{(1)} = \sum_{s}\dfrac{s\sigma_{1}}{12R_{s}^{4}}
~{\rm sgn} (R_{s})\Big[2\mathcal{A}_{x}
-6\mathcal{F}^{1/2}\left(\delta^{1}_{s}+\delta^{3}_{s}\right)
\\
+2\left(p^{3}-q^{3}\right)-9\left(p^{2}+q^{2}\right)-6\mathcal{J}\left(p-q\right)
\Big]~.
\end{multline}
Here $\mathcal{A}_{1}^{x}$ is defined earlier and $\mathcal{J}=\left(R_{s}^{2}-3\right)$. The $B$-linear correction to the off-diagonal components of the conductivity matrix is identically zero, i.e.,  $\sigma_{xy}^{(1)}=\sigma_{xz}^{(1)}=\sigma_{yz}^{(1)}=0$.

Now we calculate the quadratic-$B$ components of the conductivities. Correction to the diagonal component along the direction of the tilt is given by
\begin{multline} 
\sigma_{zz}^{(2)}=\sum_{s}\dfrac{\sigma_{0}}{4|R_{s}|^{5}}\Big[2\mathcal{A}_{2}^{z}-5\left(p^{6}+q^{6}\right)+24\left(p^{5}-q^{5}\right)
\\
+15\mathcal{J}\left(p^{4}+q^{4}\right)-~40\mathcal{F}^{1/2}\left(p^{3}-q^{3}\right)-15\mathcal{F}\left(p^{2}+q^{2}\right)\Big]~,
\end{multline}
where $\mathcal{A}_{2}^{z}=\left(1-5R_{s}^2+15R_{s}^4+5R_{s}^6\right)$. Correction to the other diagonal components is given by $\sigma_{xx}^{(2)}=\sigma_{yy}^{(2)}$, where
\begin{multline}
\sigma_{xx}^{(2)}=\sum_{s}\dfrac{\sigma_{0}}{16|R_{s}|^{5}}\Big[2\mathcal{A}_{2}^{x}+10\left(p^{6}+q^{6}\right)-48\left(p^{5}-q^{5}\right)
\\
-15\mathcal{L}\left(p^{4}+q^{4}\right)+~40\mathcal{N}\left(p^{3}-q^{3}\right)-30\mathcal{F}^{1/2}\left(p^{2}+q^{2}\right)\Big]~.
\end{multline}
%
Here we have defined 
$\mathcal{A}_{2}^{x}=\left(-2+5R_{s}^2+5R_{s}^6\right)$, $\mathcal{N}= \left(R_{s}^2-2\right)$
and $\mathcal{L}=\left( R_{s}^2-6\right)$.
Similar to the case of type-I WSMs, the quadratic corrections to the transverse conductivity components are identically zero.

\subsection{Magnetic field perpendicular to the tilt (${\bf B} \perp {\bf R}_{s}$)}
In this section we discuss transport coefficients when magnetic field is applied in a plane perpendicular to the tilt axis of the WNs. 
As expected, in this scenario there is no $B$-linear correction in the diagonal component. The quadratic corrections to the diagonal elements related to the longitudinal Hall conductivity are given by $\sigma_{yy}^{(2)}(\phi)=\sigma_{xx}^{(2)}(\pi/2-\phi)$, where $\sigma_{xx}^{(2)}=\sigma_{4}\cos^2\phi+ \sigma_{5}\sin^2 \phi$ with
\begin{multline}
\sigma_{4}=\sum_{s}\dfrac{\sigma_{0}}{32R^5}\Big[2\mathcal{B}_{2}^{c}
-15\left(p^{6}+q^{6}\right)-24\mathcal{X}\left(p^{5}-q^{5}\right)
\\
-15\mathcal{Y}\left(p^{4}+q^{4}\right)
+40\mathcal{Z}\left(p^{3}-q^{3}\right)-45\mathcal{F}\left(p^{2}+q^{2}\right)\Big]~.
\end{multline}
%
The above expression is written in terms of the following polynomials of $R_s$:
\begin{eqnarray}\nn
\mathcal{B}_{2}^{c}&=&3-7R_{s}^2+25R_{s}^4+255R_{s}^6+60R_{s}^8~,\\\nn
\mathcal{Y}&=& 9-11R_{s}^2+4R_{s}^{2}~,\\\nn
\mathcal{Z}&=& 3-5R_{s}^2+2R_{s}^4~,
\end{eqnarray}
and $\mathcal{X}=2R_{s}^2-3$. The coefficient of $\sin^2\phi$, which originates from the phase-space factor is given by
\begin{multline}
\sigma_{5}=\sum_{s}\dfrac{\sigma_{0}}{32|R_{s}|^5}\Big[2\mathcal{B}_{2}^{s}-5\left(p^{6}+q^{6}\right)
+24\left(p^{5}-q^{5}\right)
\\
+15\mathcal{J}\left(p^{4}+q^{4}\right)-40\mathcal{F}^{1/2}\left(p^{3}-q^{3}\right)-15\mathcal{F}\left(p^{2}+q^{2}\right)\Big]~.
\end{multline}
Here, $\mathcal{B}_{2}^{s}=\left(1-5R_{s}^2+15R_{s}^4+5R_{s}^6\right)$. 
The quadratic-$B$ component of the other longitudinal conductivity is given by
\begin{multline}
\sigma_{zz}^{(2)}=\sum_{s}\dfrac{\sigma_{0}}{16|R_{s}|^5}\Big[2\mathcal{C}_{2}+10\left(p^{6}+q^{6}\right)
+24\mathcal{N}\left(p^{5}-q^{5}\right)
\\
+15\mathcal{V}\left(p^{4}+q^{4}\right)-80\mathcal{F}\left(p^{3}-q^{3}\right)-30\mathcal{F}^{3/2}\left(p^{2}+q^{2}\right)\Big]~.
\end{multline}
Here, we have used the following polynomials of $R_s$: 
\begin{eqnarray}\nn
\mathcal{P}_{2}&=&-2+11R_{s}^2-25R_{s}^4+65R_{s}^6+15R_{s}^8~,\\\nn
\mathcal{V}&=&6-7R_{s}^2+R_{s}^4~.
\end{eqnarray}

The planar Hall conductivity is quadratic in $B$ and it is given by 
\begin{multline}
\sigma_{xy}^{(2)}=\sum_{s}\dfrac{\sigma_{0}}{32|R_{s}|^5}\Big[2\mathcal{D}_{2}-5\left(p^{6}+q^{6}\right)
-24\mathcal{F}^{1/2}\left(p^{5}-q^{5}\right)
\\
-15\mathcal{Z}\left(p^{4}+q^{4}\right)+40\mathcal{F}\left(p^{3}-q^{3}\right)-15\mathcal{F}\left(p^{2}+q^{2}\right)\Big]\sin2\phi~.
\end{multline}
Here we have defined
\begin{eqnarray}\nn
\mathcal{D}_{2}&=&1-R_{s}^2+5R_{s}^4+125R_{s}^6+30R_{s}^8~.
\end{eqnarray}
Unlike the in plane transverse components, the out-of-plane transverse components have only linear-$B$ correction and they can be written as $\sigma_{yz}^{(1)}=\sigma_{3}\sin\phi$ and $\sigma_{xz}^{(1)}=\sigma_{3}\cos\phi$, where 
\begin{multline}
\sigma_{3}=\sum_{s}\dfrac{s\sigma_{1}}{12R_{s}^4}~{\rm sgn}(R_{s})\Big[2\mathcal{M}+6\mathcal{F}(\delta_{s}^{1}+\delta_{s}^{3})
\\
+2\left(p^{3} - q^{3}\right)+9\mathcal{F}^{1/2}\left( p^2 + q^{2}\right)+6\mathcal{G}\left(p-q\right)\Big]~.
\end{multline}}

\section{Lorentz symmetry breaking of tilted Dirac and Weyl nodes}
To start with, we prove the Lorentz invariance of Dirac equation without any tilt. Then, we show that adding a tilt term such as 
$\gamma^{0}\partial_{z}$ to the massless Dirac Hamiltonian,  breaks the Lorentz invariance. 

The coordinates' transformation under a Lorentz boost is given by ${x^{\prime}}^{\mu}=\Lambda_{\nu}^{\mu}x^{\nu}$.  Accordingly, the derivative transform as $ \partial_i \to (\Lambda^{-1})^{\nu}_{i}\partial_{\nu}$. 
Recall that the Dirac equation is given by
\be
[i\gamma^{\mu}\partial_{\mu}-m]\Psi(x)=0~,
\ee
with $\gamma^{\mu}$ denoting the $4\times4$ Dirac gamma matrices and $\mu=\{0,1,2,3\}$.
In the boosted frame, the left-hand side of the above equation takes the form $[i\gamma^{\mu}\partial_{\mu}^{\prime}-m]\Psi^{\prime}(x^{\prime})$. 
The Lorentz invariance of the Dirac equation requires transformation of spinor as $\Psi^{\prime}(x^{\prime})=S\Psi(x)$, where the transformation matrix satisfies
\begin{equation}
S^{-1}\gamma^{\mu}S=\Lambda_{\nu}^{\mu}\gamma^{\nu}.
\end{equation}

Now, in condensed matter systems the Weyl fermions can in principle occur in tilted WNs whose dispersion is well captured by the massless tilted Dirac equation, 
\begin{equation} \label{eq_Weyl}
i[\gamma^{\mu}\partial_{\mu} + \gamma^{0}\partial_{z} ]\Psi(x)=0~.
\end{equation}
This describes WNs with tilt in the $z$ direction.  
Note that the first term of this equation is Lorentz invariant \cite{Peskin05}. But the last term transforms to
\begin{eqnarray}\nonumber
i\gamma^{0}\partial_{z}^{\prime}\Psi^{\prime}(x^{\prime})&= &\left[i\gamma^{0}(\Lambda^{-1})^{\nu}_{z}\partial_{\nu}\right]S\Psi( x)
\\\nonumber
&=&S S^{-1}\left[i\gamma^{0}(\Lambda^{-1})^{\nu}_{z}\partial_{\nu}\right]S\Psi( x)
\\\nonumber
&=&S\left[iS^{-1}\gamma^{0}S(\Lambda^{-1})^{\nu}_{z}\partial_{\nu}\right]\Psi( x)
\\\nonumber
&=&S\left[i\Lambda^{0}_{\alpha}\gamma^{\alpha}(\Lambda^{-1})^{\nu}_{z}\partial_{\nu}\right]\Psi( x)
\\\nonumber
&=&S\left[i\gamma^{\alpha}\Lambda^{0}_{\alpha}(\Lambda^{-1})^{\nu}_{z}\partial_{\nu}\right]\Psi( x)~.
\end{eqnarray}
For a particular choice of $\Lambda$, we can show that $\gamma^{\alpha}\Lambda_{\alpha}^{0}(\Lambda^{-1})_{z}^{\nu}\partial_{\nu}=\Gamma(\gamma^{0}-\beta\gamma^{1})\partial_{z}$, where $\Gamma=(1-\beta^2)^{-1/2}$ and $\beta={v}/{c}$.
Thus this term does not retain the original form, leading to the breakdown of the Lorentz invariance in tilted WSMs.

\section{Energy dependence of $\tau_\mu$}
\label{secF}
Here, we discuss the energy dependence of the intranode scattering timescale in a WN \cite{Burkov11,Das_sharma15}.  
More specifically, we consider the chemical potential dependence of $\tau_\mu$ arising from the short-ranged disorder (neutral point defects), and charged impurities. 

Modelling the neutral point defects as a delta-function impurity, and using the Born approximation, the scattering timescale $\tau_{\delta}(\epsilon)$, has been 
calculated \cite{Burkov11,Das_sharma15} and it is given by 
\begin{equation}\label{neutral_impurity}
\dfrac{1}{\tau_{\delta}(\epsilon)}=\dfrac{n_{i}V_{0}^2\epsilon^2}{3\pi \hbar^4 v_{F}^3}~.
\end{equation}
Here, $V_{0}$ is the strength of the delta-function potential and $n_{i}$ is the impurity density, and $\epsilon$ is the energy scale 
which typically corresponds to the chemical potential.  

Next we consider screened Coulomb disorder of the charged impurities. Considering a Thomas-Fermi screened Coulomb potential, the energy dependence of a 
screened Coulomb impurity was determined to be \cite{Das_sharma15} 
\begin{equation}
\dfrac{1}{\tau_{c}}= 4\pi n_{i}\alpha^2\dfrac{\hbar^2v_{F}^3}{\epsilon^2}~I_{t}(q_{0})~.
\end{equation}
Here, $\alpha=e^2/\kappa\hbar v_{F}$ is the effective coupling constant with dielectric constant $\kappa$, and $q_{0}=q_{TF}/2k_{F} = \sqrt{\alpha/2\pi}$ in a 
Weyl node. In addition, we have \cite{Das_sharma15}
\begin{equation}
I_{t}(q_{0})=\left(q_{0}^{2}+\frac{1}{2}\right)\ln\left(1+\dfrac{1}{q_{0}^{2}}\right)-1~.
\end{equation}
The chemical potential dependence of the effective intranode scattering scale can be obtained by combining both these timescales,  
\begin{equation}
\frac{1}{\tau_\mu} = \frac{1}{\tau_\delta} + \frac{1}{\tau_{c}}~.
\end{equation}

Note that in principle for anisotropic Weyl nodes, $\tau_\mu$ may also be anisotropic. However, for simplicity we will ignore the anisotropy 
of $\tau_\mu$ in this paper. 
}

\bibliography{LMR_WSM_v1.bib}
\end{document}